\documentclass[final,authoryear,3p,times]{elsarticle}
\usepackage{graphicx,color,colortbl}
\usepackage{enumerate}
\usepackage[colorlinks]{hyperref}
\usepackage{amsmath,amssymb,amsthm,bm}
\usepackage{multicol}
\usepackage[dvipsnames]{xcolor}
\usepackage{ulem}
\usepackage{tikz}
\usetikzlibrary{automata,positioning}
\usepackage{afterpage}
\usepackage{calc}% http://ctan.org/pkg/calc
\usepackage{ifthen}
\usepackage[ruled, vlined, linesnumbered]{algorithm2e}
\usepackage{algpseudocode}
\usepackage{caption}
\usepackage{subcaption}
\usepackage{siunitx}
\usepackage{booktabs}
\usepackage{placeins}
\usepackage{float}

\usepackage{multirow}

\edef\svtheparindent{\the\parindent}
\usepackage{parskip}
\parindent=\svtheparindent\relax

\newlength{\boxwidth}
\def\btheorem{\begin{theorem}}
\def\etheorem{\end{theorem}}
\def\blemma{\begin{lemma}}
\def\elemma{\end{lemma}}
\def\bproposition{\begin{proposition}}
\def\eproposition{\end{proposition}}
\def\bcorollary{\begin{corollary}}
\def\ecorollary{\end{corollary}}
\def\bdefinition{\begin{definition}}
\def\edefinition{\end{definition}}
\def\bexample{\begin{example}}
\def\eexample{\end{example}}
\def\bremark{\begin{remark}}
\def\eremark{\end{remark}}

\newcommand{\be}{\begin{equation}}
\newcommand{\ee}{\end{equation}}
\newcommand{\beq}{\begin{eqnarray}}
\newcommand{\eeq}{\end{eqnarray}}
\newcommand{\bem}{\begin{multline}}
\newcommand{\eem}{\end{multline}}
\newcommand{\ba}{\begin{align}}
\newcommand{\ea}{\end{align}}

\journal{CMAME}

\begin{document}

\begin{frontmatter}

\title{Floating Isogeometric Analysis}

\author[eth]{Helge C. Hille}
\author[tud]{Siddhant Kumar}
\author[eth]{Laura De Lorenzis\corref{cor1}}

\cortext[cor1]{Correspondence: ldelorenzis@ethz.ch}

\address[eth]{Department of Mechanical and Process Engineering, ETH Z\"{u}rich, 8092 Z\"{u}rich, Switzerland}
\address[tud]{Department of Materials Science and Engineering, Delft University of Technology, 2628 CD Delft, The Netherlands}

\begin{abstract}
We propose Floating Isogeometric Analysis (FLIGA), which extends the concepts of IGA to Lagrangian extreme deformation analysis. The method is based on a novel tensor-product construction of B-Splines for the update of the basis functions in one direction of the parametric space. With basis functions ``floating'' deformation-dependently in this direction, mesh distortion is overcome for problems in which extreme deformations occur predominantly along the associated (possibly curved) physical axis. In doing so, we preserve the numerical advantages of splines over many meshless basis functions, while avoiding remeshing. We employ material point integration for numerical quadrature attributing a Lagrangian character to our technique. The paper introduces the method and reviews the fundamental properties of the FLIGA basis functions, including a numerical patch test. The performance of FLIGA is then numerically investigated on the benchmark of Newtonian and viscoelastic Taylor-Couette flow. Finally, we simulate a viscoelastic extrusion-based additive manufacturing process, which served as the original motivation for the new approach.
	
\end{abstract}


\end{frontmatter}

%%%%%%%%%%%%%%%%%%%%%%%%%%%%%%%%%%%%%%%%%%%%%%%%%%%%%%%%%%%%%%%%%%%%%%%%%%%%%%%%%%%%%%%%%%%%%%%%%%%%%%%%%%%%%%%%%%%%%%%

\section{Introduction}\label{sec:Introduction}

Isogeometric analysis (IGA) is a generalization of finite element analysis (FEA) which adopts basis functions from the geometry representation in Computer Aided Design (CAD) instead of the standard $\mathcal{C}^0$-continuous piecewise polynomials. The first isogeometric techniques were proposed by \cite{HUGHES20054135} with the aim to naturally bridge the gap between CAD and analysis. Since then and beyond the original goal, isogeometric basis functions proved to lead to many advantages for the analysis itself, most notably in relation to their higher and tailorable continuity.

One of the favorable features of IGA is a reduced sensitivity to mesh distortion compared to FEA \citep{LIPTON2010357}. Nevertheless, beyond a certain deformation limit, standard (Lagrangian) IGA techniques also suffer from mesh distortion. This issue is common to all mesh-based methods and manifests itself with decreasing result accuracy and final loss of solvability under increasing material deformation.

A common technique to overcome mesh distortion in FEA is remeshing. In such approaches, before critical entanglement occurs, the original mesh is replaced by a new one and the needed information is mapped from the old to the new mesh. Remeshing strategies are computationally expensive and known to accumulate errors whose prediction and control is yet another challenge \citep{WIECKOWSKI20044417}. What additionally hinders remeshing from extensive application in IGA is its methodical complexity \citep{Fueder2015IsogeometricFE}. Nonetheless, research on isogeometric remeshing techniques is currently ongoing, see e.g. \cite{Shamanskiy2020}.

Another alternative to handle extreme deformations in a Lagrangian context are meshless methods, which avoid mesh distortion by giving up the mesh altogether \citep{BELYTSCHKO19963,Chen2017}. In these methods, the basis functions are computed recurrently from scattered nodal points such that distortion is avoided, and thus follow material deformations in a loosened sense. Note that only the Lagrangian character of the basis functions is abolished, but that of the evaluation points as well as of the governing equations is preserved, so that many meshless methods may still be considered Lagrangian techniques. Unfortunately, the isogeometric goal of a unification of CAD and analysis is not compatible with the meshless approach. Moreover, meshless methods are frequently less efficient than their mesh-based competitors \citep{XIAO2020,Kumar2020nme} and suffer from well-known instabilities \citep{BELYTSCHKO2000,KUMAR2019858}. Moreover, for the case of incompressible material behavior which is relevant in this paper, the construction of mixed discretizations to avoid locking is not trivial \citep{GOH2018575}. The efficiency and stability of mesh-based computations and the flexibility of meshless methods have motivated some hybrid approaches, aiming to benefit from the complementing advantages of both settings, see e.g. \cite{ArroyoOrtiz2006,ROSOLEN201395, MILLAN2015712, CARDOSO2014, FATHI2021}.

A third prominent way to handle extreme deformations is the change from the Lagrangian to the Eulerian viewpoint, which has been widely adopted also within the isogeometric paradigm for fluid mechanics applications \citep{AKKERMAN20114137,Evans2013,HOSSEINI2015264,HOSSEINI2017171,
NIELSEN20113242}. However, these approaches typical of computational fluid dynamics (CFD) are not optimal for problems involving solid material behavior such as viscoelasticity or plasticity, which are the focus in this paper. Moreover, Eulerian formulations typically come along with increased numerical complexity since a stabilization of the advective terms as well as a special treatment of moving boundaries are required, see e.g. \cite{AKKERMAN20114137}. 

In this paper, we propose a hybrid method that adopts meshless concepts to generalize mesh-based IGA. We denote our method as \textit{floating} IGA (FLIGA) and specifically design it for mechanics problems where it is possible to idenfity a single (possibly curved) direction of predominant deformation, which we call \textit{characteristic direction}. Moderate deformations may still occur perpendicular to this direction. The proposed method is based on the idea of introducing meshless behavior in IGA only as far as the isogeometric concept is preserved, but to such an extent that mesh distortion is overcome. We realize this by breaking the Lagrangian character of isogeometric basis functions and preventing their distortion along the characteristic direction. Yet, the Lagrangian character of governing equations and quadrature points is preserved.

\begin{figure}[h!]
\centering
\includegraphics[trim={4cm 1.75cm 4cm 1.75cm},clip,width=13cm]{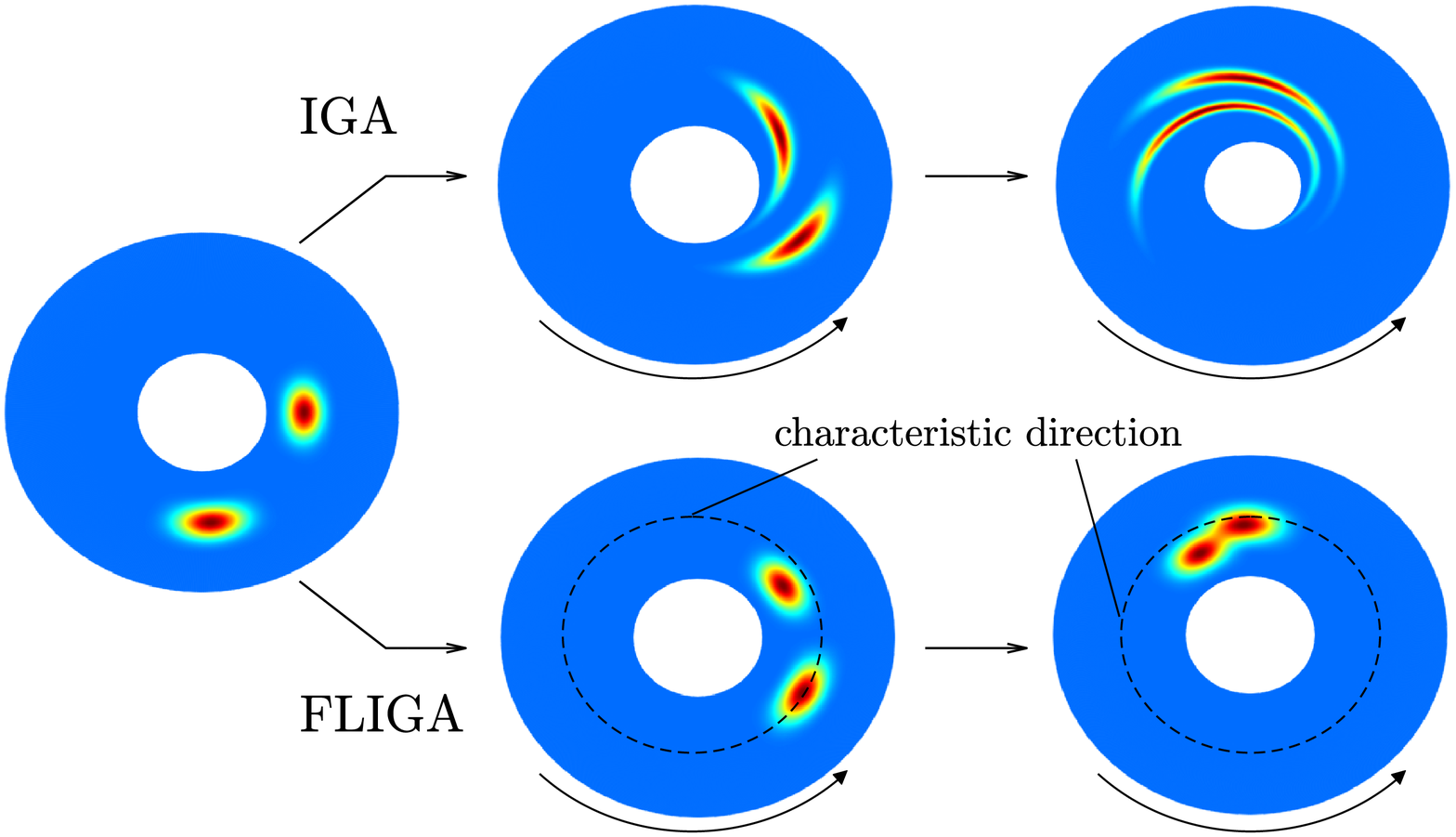}
\caption{Comparison of basis function evolution for IGA and FLIGA in the Taylor-Couette flow problem. In IGA, basis functions strictly follow the deformation of the material leading to severe distortion. In FLIGA, we can determine a characteristic direction, where basis functions follow the deformation only in an average sense.}
\label{fig:fligadistortion}
\vspace*{1cm}
\includegraphics[width=12cm]{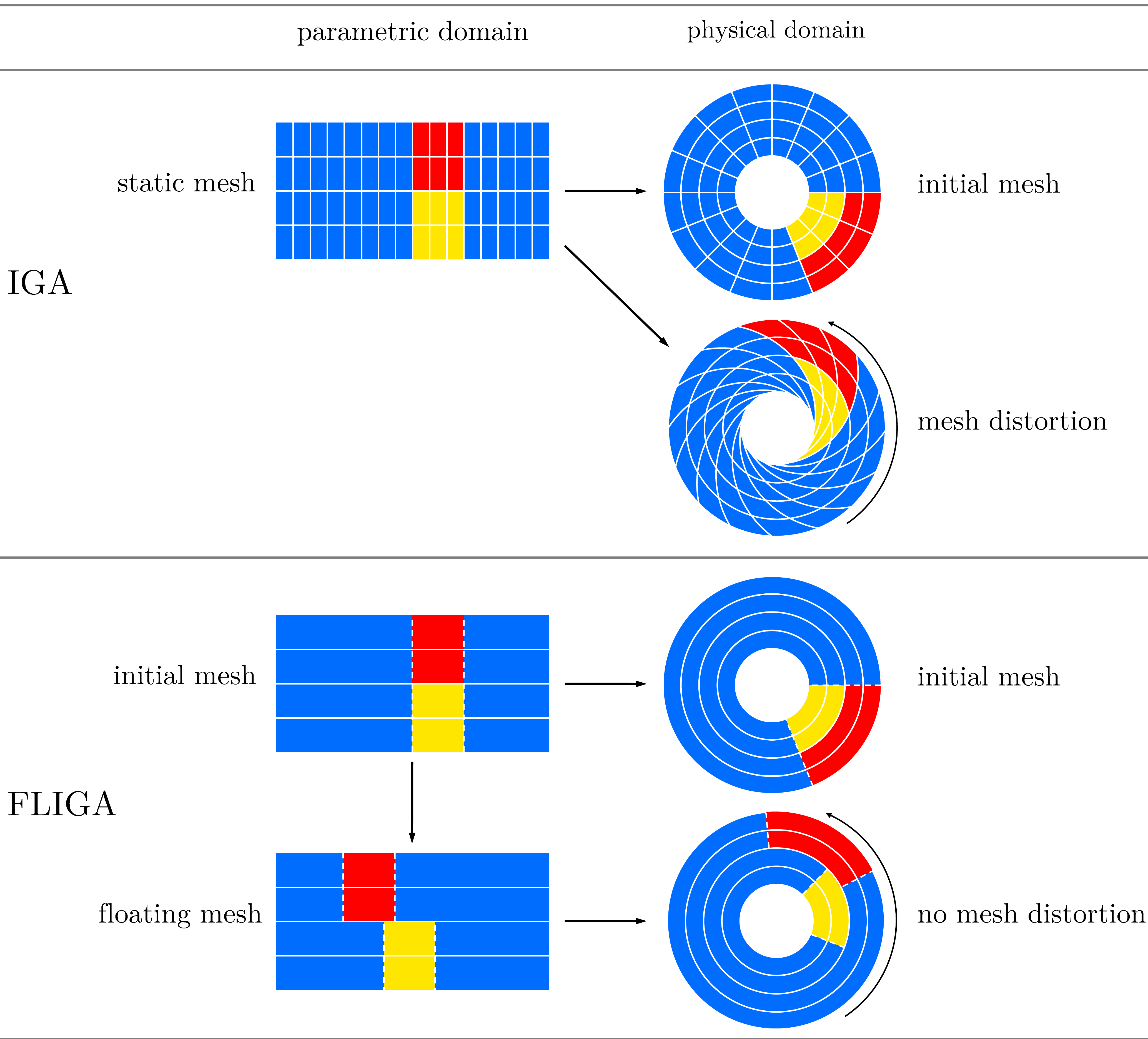}
\caption{Evolution of the support of two different basis functions (colored in yellow and red) in the parametric and physical domains for IGA and FLIGA in the Taylor-Couette flow problem.}
\label{fig:fligaconcept3}
\end{figure}

The behavior of FLIGA basis functions is qualitatively illustrated in \autoref{fig:fligadistortion}, which refers to the Taylor-Couette flow with the internal boundary fixed and the external one rotating counterclockwise. Here the characteristic direction of deformation is the circumferential one. Through the floating procedure, the two illustrated basis functions are prevented from undergoing extreme shear deformations.
\autoref{fig:fligaconcept3} shows schematically how this behavior is achieved. We start from a classical IGA discretization where one of the straight axes of the rectangular parametric domain (the horizontal one in the figure) is mapped to the characteristic (circumferential) direction in the physical space. We modify the classical tensor-product structure of multivariate B-Spline bases in the parametric domain, in that we grant each basis function the ability to float along the characteristic direction independently from its neighbors in the normal direction. This floating is carried out depending on the deformation and allows to significantly reduce the basis function distortion after finally mapping the parametric domain to physical space. Note that this comes along with a loosening of the classical mesh notion, as the element boundaries perpendicular to the characteristic direction are no longer in place, whereas those along the characteristic direction are kept.

\begin{figure}[bt]
\centering
\includegraphics[trim={4cm 1.75cm 3.5cm 1.75cm},clip,width=13cm]{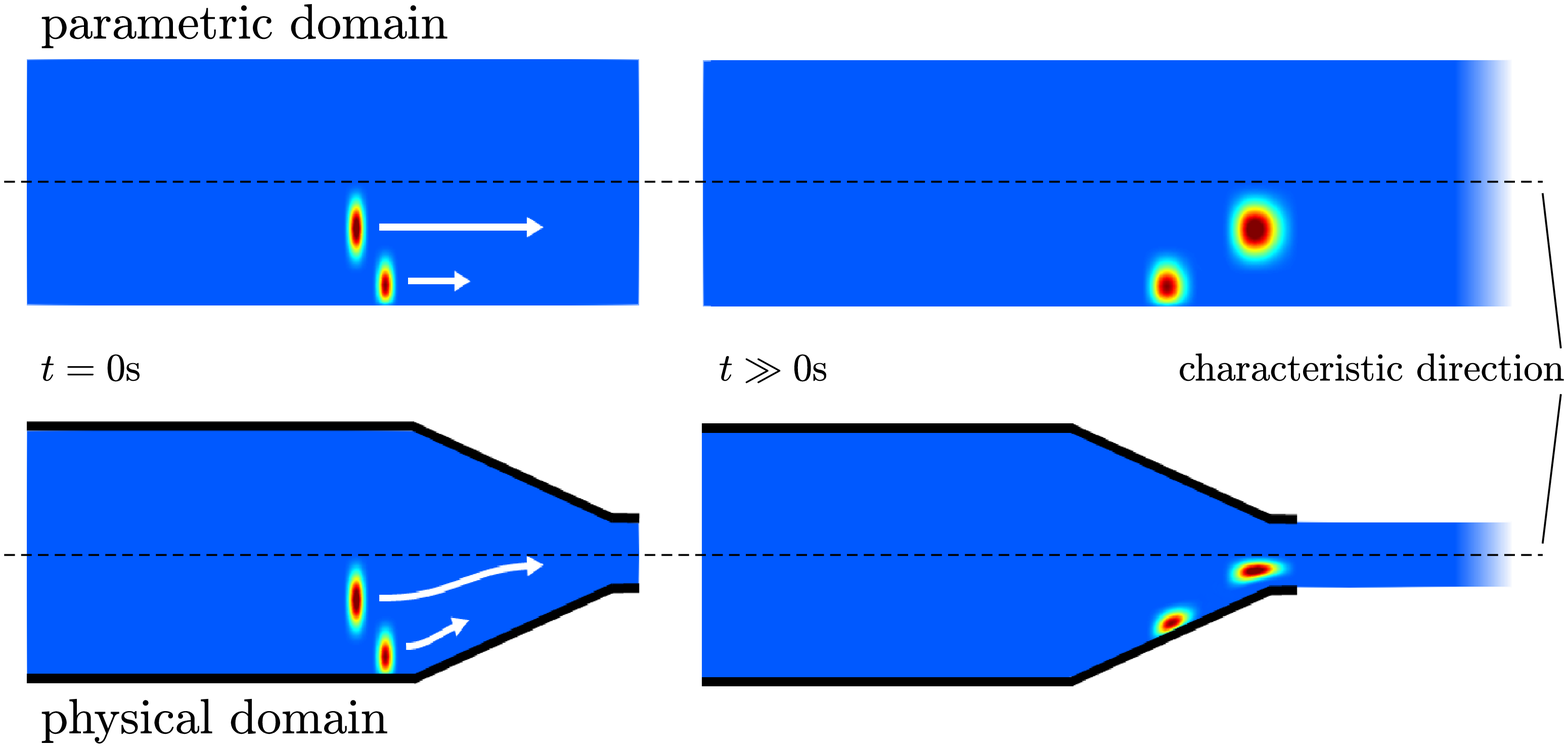}
\caption{Schematic basis function evolution for FLIGA in the example of frictional extrusion.}
\label{fig:fligaextrusiondistortion}
\end{figure}

Our development of FLIGA is motivated by the simulation of polymer extrusion, one of the main additive manufacturing (AM) processes, in which extreme deformations occur along the extrusion direction (\autoref{fig:fligaextrusiondistortion}). The investigations conducted so far made use of remeshing, meshless methods, or CFD techniques. E.g., \cite{REINOLD2020} investigated the formation of layer geometries and the layer interaction within large-scale extrusion by the Particle Finite Element Method, a remeshing technique based on FEA. Smoothed Particle Hydrodynamics was adopted in \cite{OUYANG2020} to study fused deposition modeling of a fiber-reinforced polymer, with special attention to the alignment of fibers in the viscous melt. \cite{COMMINAL2018} employed CFD simulations to predict viscous strand deposition flow for small-scale AM. 

We believe that FLIGA bears great potential for simulation of extrusion-based AM. Due to the Lagrangian character of FLIGA, we easily treat the viscoelastic behavior of polymers without the need to stabilize advective terms. The meshing efforts are minimized, as the geometry of the nozzle may be obtained seamlessly from its CAD representation. The contact between nozzle and polymer melt is treated robustly due to the smoothness of the basis functions, see  \cite{DeLorenzis2014}. Most importantly, despite the extreme deformations, FLIGA does not suffer from mesh distortion avoiding the drawbacks of remeshing. Free surfaces that form behind the nozzle exit are naturally handled without additional efforts to track boundaries. Incompressible material behavior can be treated by well-established mixed approximation techniques from mesh-based IGA and no stabilization is required. Finally, by staying within the solid mechanics setting we are ideally prepared to study, e.g., the evolution of the polymer melt behavior and properties upon cooling, residual stresses and final part properties.

The outline of the paper is as follows: In Section \ref{sec:IGA} we review our continuum model and its discretization with IGA, which are the basis for the introduction of FLIGA in Section \ref{sec:FLIGA}. Numerical examples are provided in Section \ref{sec:results} and conclusions drawn in Section \ref{sec:conclusions}.

\section{Continuum model and isogeometric analysis}
\label{sec:IGA}

In this section, we briefly recall the fundamental equations of solid mechanics for incompressible materials and review their isogeometric discretization. The formulation is based on velocities rather than on displacements, as it is customary in the literature on viscous flow. For simplicity, in this paper we limit ourselves to the two-dimensional case, however the extension to three dimensions is conceptually straightforward.

\subsection{Continuum equations}
\label{ssec:continuumequations}

The basis for our discretization is the Principle of Virtual Power
\begin{equation}
\int_\Omega \bm{\sigma} : \delta \bm{D}\ \mathrm{d\Omega} = \int_{\partial\Omega_N} \bm{h}_N \cdot \delta \bm{v}\ \mathrm{d\Gamma}
\label{eq:PVWcontinuous}
\end{equation}
where for simplicity we neglect inertia and body force contributions. Here, $\bm{\sigma}$ denotes the Cauchy stress, $\bm{D}$ is the strain rate tensor related to the velocity $\bm{v}$ by 

\begin{equation}
\bm{D} = \dfrac{1}{2}\left(\nabla_{\bm{x}}\bm{v}+\nabla^T_{\bm{x}}\bm{v}\right) 
\label{eq:kinematicscontinuous}
\end{equation}
and $\delta\bm{D}$ and $\delta\bm{v}$ are respectively the virtual strain rate and the virtual velocity, also related by
\begin{equation}
\delta\bm{D} = \dfrac{1}{2}\left(\nabla_{\bm{x}}\delta\bm{v}+\nabla^T_{\bm{x}}\delta\bm{v}\right)
\label{eq:kinematicscontinuous}
\end{equation}
We are using the notation $\left(\nabla_{\bm{x}}(\bullet)\right)_{ij}={\partial (\bullet)_i}/{\partial x_j}$ to denote the spatial gradient of a first-order tensor $(\bullet)$, and $\bm{x}\in\Omega$ is the coordinate of a material point in the current (deformed) configuration. The boundary of the spatial domain $\Omega$, $\partial\Omega$, consists of two subsets $\partial\Omega_D$ and $\partial\Omega_N$, with $\partial\Omega_D \cup \partial\Omega_N = \partial \Omega$ and $\partial\Omega_D \cap \partial\Omega_N = \emptyset$. On the Dirichlet boundary the velocity is prescribed
\begin{equation}
\bm{v}=\bm{v}_D \ \ \ \ \ \ \ \ \forall\ \bm{x} \in \partial\Omega_D
\label{eq:Dir}
\end{equation}
whereas the external surface force vector $\bm{h}_N$ acts on the Neumann boundary 
\begin{equation}
\bm{\sigma}\cdot\bm{n}=\bm{h}_N \ \ \ \ \ \ \ \ \forall\ \bm{x} \in \partial\Omega_N
\label{eq:Neu}
\end{equation}
where $\bm{n}$ is the outward normal unit vector at each point of the boundary. 
%With the Principle of Virtual Power, we seek $\bm{v}$ sufficiently smooth and satisfying \eqref{eq:Dir}, such as \eqref{eq:PVWcontinuous} holds for each $\delta\bm{v}$ sufficiently smooth and satisfying the homogeneous counterpart of \eqref{eq:Dir}. 

Without loss of generality, we decompose the Cauchy stress $\bm{\sigma}$ in a volumetric and a deviatoric contribution: 
\begin{equation}
\bm{\sigma} = -p\bm{\mathrm{I}} + \bm{\tau}
\label{eq:constitutivesplit}
\end{equation}
where $p$ is the pressure, $\bm{\mathrm{I}}$ the second-order identity tensor and $\bm{\tau}$ the deviatoric part of the Cauchy stress.
We give more details on the constitutive relationships later, when applying the Principle of Virtual Power to specific materials with viscous and viscoelastic behavior.

In addition to the Principle of Virtual Power, we require incompressibility of the velocity field, i.e.
\begin{equation}
\nabla_{\bm{x}}\cdot\bm{v}=0 \ \ \ \ \ \ \ \ \forall\ \bm{x} \in \Omega
\end{equation}
which we recast in weak form as
\begin{equation}
\int_\Omega  \left( \nabla_{\bm{x}}\cdot\bm{v}\right)  \delta p\ \mathrm{d\Omega} = 0
\end{equation}
where $\delta p$ is a virtual scalar pressure field.

In problems where we do not impose any Neumann boundary conditions, i.e. $\partial\Omega_D=\partial\Omega$, the pressure is not uniquely determined, but allows for an arbitrary constant offset. In such cases, we achieve a unique pressure solution by enriching the problem with a Dirichlet pressure prescription $p_D$ at a suitable part of the boundary $\partial\Omega_{D,p}\in\partial\Omega$.
\begin{equation}
p=p_D \ \ \ \ \ \ \ \ \forall\ \bm{x} \in \partial\Omega_{D,p}
\end{equation}

Note that all the fields involved in the previous equations undergo spatial and temporal variation, which we have not explicitly written for notational simplicity. We can obtain the transient evolution of a material particle $\bm{x}_p$ by means of
\begin{equation}
\bm{x}_p\left(\bm{x}^0_p,t\right) = \bm{x}^0_p + \int_0^t \bm{v}_p\left(\bm{x}_p(\bm{x}^0_p,\tau),\tau\right) \mathrm{d\tau}
\label{eq:lagrangianviewpoint}
\end{equation}
where $\bm{x}^0_p$ denotes the initial particle position at time $t=0$.

\subsection{B-Splines}
\label{ssec:bsplines}

In this section, we summarize the construction of isogeometric basis functions, which are needed for the subsequent isogeometric discretization. Here we limit ourselves to B-Splines, however, in the past years, other function types from the CAD community have been employed in the context of IGA, e.g. NURBS, T-Splines, and a few more \citep{HUGHES20054135,BAZILEVS2010229}.

Given the integers $r\geq0$ and $I\geq r+1$, we define a knot vector as
\begin{equation}
\Xi=\left\lbrace \xi_{1},\xi_{2},\ldots,\xi_{i},\ldots,\xi_{I+r+1}\ \vert\ \xi_i \in \mathbb{R}\right\rbrace
\end{equation}
Here we assume repetition of the first $r+1$ and last $r+1$ knots and monotonic increase for the entire knot vector, i.e., $\xi_1=\xi_{r+1}$ and $\xi_{I+1}=\xi_{I+r+1}$, while $\xi_{i} \leq \xi_{i+1}$. Note that we will assume the same for all other knot vectors introduced later, also beyond this section.

Let us now span a parametric domain $\hat{\Omega}_{\xi}=\left[\xi_1, \xi_{I+r+1}\right]$ over coordinate $\xi$. A univariate (denoted for brevity as 1D) B-Spline basis $\left\lbrace\hat{N}_{i,r}\left(\xi\right)\right\rbrace^{I}_{i=1}$ on $\hat{\Omega}_{\xi}$ with polynomial order $r$ can be constructed by the Cox-de Boor recursion formula
\begin{equation}
\begin{split}
r=0:\ \ \ \hat{N}_{i,0}\left(\xi\right)&= \left\{
\begin{array}{ll}
1 & \textrm{for}\  \xi_{i} \leq \xi < \xi_{i+1} \\
0 & \, \textrm{otherwise} \\
\end{array}
\right.\\
r\geq 1:\ \ \ \hat{N}_{i,r}\left(\xi\right)&=\frac{\xi-\xi_i}{\xi_{i+r}-\xi_i}\hat{N}_{i,r-1}\left(\xi\right)+\frac{\xi_{i+r+1}-\xi}{\xi_{i+r+1}-\xi_{i+1}}\hat{N}_{i+1,r-1}\left(\xi\right)
\end{split}
\label{eq:coxdeboor}
\end{equation}
where $\xi_i \in \Xi$. It holds $\hat{N}_{i,r-1}=0$, if $\xi_i=\xi_{i+r}$; as well as $\frac{0}{0}=0$.
For more concise notation, from now on we will omit the polynomial order index $r$ when referring to B-Spline bases or their individual functions. 

\begin{figure}[t]
\centering
\includegraphics[trim={6cm 2cm 3.5cm 9.75cm},clip ,width=13.5cm]{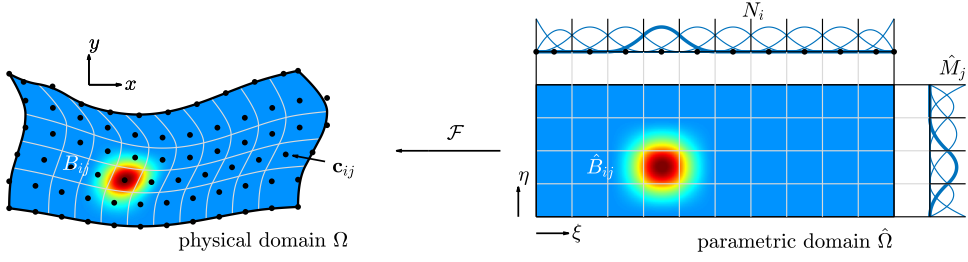}
\caption{Construction of classical bivariate tensor-product B-Splines.}
\label{fig:tensorproduct}
\end{figure}

In classical IGA, a bivariate (also denoted for brevity as 2D) B-Spline basis $\left\lbrace\hat{B}_{ij}\left(\bm{\xi}\right)\right\rbrace^{I,J}_{i=1,j=1}$ is constructed by means of a tensor-product structure.
\begin{equation}
\hat{B}_{ij}\left(\xi,\eta\right)=\hat{N}_{i}\left(\xi\right)\hat{M}_{j}\left(\eta\right)
\label{eq:tensorproductstructure}
\end{equation}
in which $\hat{M}_{j}\in \left\lbrace\hat{M}_{j}\left(\eta\right)\right\rbrace^{J}_{j=1}$ are again 1D B-Spline basis functions of polynomial order $r$, however, defined on $\hat{\Omega}_{\eta}=[\eta_1,\eta_{J+r+1}]$ and associated with knot vector $\mathcal{H}=\left\lbrace \eta_{1},\eta_{2},\ldots,\eta_{j},\ldots,\eta_{J+r+1}\ \vert\ \eta_j \in \mathbb{R} \right\rbrace$. Hence, the 2D B-Spline basis consists of $I\cdot J$ basis functions that span the parametric domain $\hat{\Omega}=\hat{\Omega}_{\xi}\times\hat{\Omega}_{\eta}$.

For the representation of B-Spline freeform surfaces we employ a linear combination $\mathcal{F}^n: \hat{\Omega}\rightarrow \Omega^n$, mapping the static rectangular parametric domain to the current physical domain
\begin{equation}
\bm{x} =\mathcal{F}^n\left(\bm{\xi}\right)= \sum_{i}\sum_{j} \bm{c}^n_{ij}\hat{B}_{ij}\left(\bm{\xi}\right)
\label{eq:Fmapping}
\end{equation}
The 2D basis can thus be expressed in terms of physical (current) coordinates as
\begin{equation}
B^n_{ij}\left(\bm{x}\right) = \hat{B}_{ij}\left(\mathcal{F}^{n^{-1}}\left(\bm{x}\right)\right)
\label{eq:PhysicalB}
\end{equation}
The linear combination coefficients $\bm{c}^n_{ij}$ in \eqref{eq:Fmapping} are the (current) coordinates of the so-called control points and are subject to transient evolution, whereas the B-Spline basis in terms of parametric coordinates does not change over time. We use the Jacobian $\bm{J}^n$ to associate parametric and physical derivatives.
\begin{equation}
\bm{J}^n\left(\bm{\xi}\right)=\nabla_{\bm{\xi}}\bm{x}=\nabla_{\bm{\xi}}\mathcal{F}^n=\sum_{i}\sum_{j}\bm{c}^n_{ij} \nabla^T_{\bm{\xi}} \hat{B}_{ij}\left(\bm{\xi}\right)
\label{eq:Fjacobian}
\end{equation}
For convenience, we summarize the indices $i,j$ in the single running index $b$, which is uniquely associated to them through
\begin{equation}
b=(j-1)\cdot I + i
\end{equation}

We summarize the above concepts graphically in \autoref{fig:tensorproduct}.
The extension to the construction of trivariate B-Spline bases and freeform volumes can be found in \cite{COTTRELL200969}.

\subsection{Discrete equations}
\label{ssec:discreteequations}

With the B-Spline basis functions at hand, we can now replace the continuum model of Section \ref{ssec:continuumequations} by a discrete formulation, limiting ourselves to the search for approximate solutions. In the literature, discretized quantities are sometimes denoted with a superscript $h$. However, we drop such an index as computational quantities are from now on always related to the discrete approximate model, unless specified otherwise. We introduce the following isogeometric field approximations for velocity $\bm{v}^n$, pressure $p^n$, virtual velocity $\delta\bm{v}^n$ and virtual pressure $\delta p^{n}$:
\begin{equation}
\bm{v}^{n}\left(\bm{x}\right) =  \sum_{b=1}^B \bm{d}^n_b B^n_b\left(\bm{x}\right)\ \ \ \ \ \                
\delta\bm{v}^{n}\left(\bm{x}\right) = \sum_{b=1}^B \delta\bm{d}^n_b B^n_b\left(\bm{x}\right) \ \ \ \ \ \ 
p^{n}\left(\bm{x}\right) = \sum_{a=1}^A q^n_a A^n_a\left(\bm{x}\right) \ \ \ \ \ \ 
\delta p^{n}\left(\bm{x}\right) = \sum_{a=1}^A \delta q^n_a A^n_a\left(\bm{x}\right)
\label{eq:ansatzes}
\end{equation}
Thus, our bases comprise $B$ B-Spline basis functions $B^n_b\left(\bm{x}\right)$ for the velocity approximation and $A$ B-Spline basis functions $A^n_a\left(\bm{x}\right)$ for the pressure approximation, and $\bm{d}^n_b$ and $q^n_a$ are the control point velocities and pressures, respectively. The virtual quantities are interpreted accordingly. We also discretize in time, which is indicated by the superscript $n$ referring to the discrete time step $t^n$. 

With these approximations, we can write the discrete form of the Principle of Virtual Power, where we also incorporate the volumetric-deviatoric split of the Cauchy stress and assume $\bm{h}_N=\bm{0}$, as follows
\begin{equation}
\int_{\Omega^n} \left(-p^n\cdot\bm{\mathrm{I}}+\bm{\tau}^n\right) : \delta \bm{D}^n\ \mathrm{d\Omega}^n = 0 
\label{eq:PVWdiscrete}
\end{equation}
Here, $\bm{\tau}^n$ denotes the discretized deviatoric Cauchy stress. We obtain the discretized version of the kinematics by replacing the continuous velocity fields in \eqref{eq:kinematicscontinuous} by the respective ansatzes from \eqref{eq:ansatzes}.

The incompressibility equation in discrete form reads
\begin{equation}
\int_{\Omega^n}  \left( \nabla_{\bm{x}}\cdot \bm{v}^n\right)  \delta p^n\ \mathrm{d\Omega}^n = 0
\label{eq:incompressibilitydiscrete}
\end{equation}

If we rewrite \eqref{eq:PVWdiscrete} by inserting the ansatz for the discretized virtual strain rate and exploiting the assumption of symmetry for the deviatoric stress tensor, for each control point index $b$ we obtain a global residual vector $\bm{S}^n_b$ that has to vanish
\begin{equation}
\bm{S}^n_b = \int_{\Omega^n} \left(-p^n\cdot\bm{\mathrm{I}}+\bm{\tau}^n\right) \nabla_{\bm{x}}B^n_b \ \mathrm{d\Omega}^n = \bm{0} \ \ \ \ \ \ \ \ \forall\ b=1,...,B
\label{eq:globalsystem1}
\end{equation}
From the weak form of the incompressibility condition we obtain another global residual vector, which again must vanish for each control point index $a$
\begin{equation}
Q^n_a = \int_{\Omega^n} \left( \nabla_{\bm{x}}\cdot \bm{v}^n\right) A^n_a \ \mathrm{d\Omega}^n = 0 \ \ \ \ \ \ \ \ \forall \ a=1,...,A
\label{eq:globalsystem2}
\end{equation}
The expressions \eqref{eq:globalsystem1} and \eqref{eq:globalsystem2} represent a global, potentially non-linear system of equations. In order for the system to be solvable we need to incorporate the discretized Dirichlet boundary conditions, which we briefly discuss in Section \ref{ssec:igacomments}. The non-linear system is commonly solved with the Newton-Raphson strategy, see \ref{globalNewtonRaphson} for more details, and the solution delivers the current control point velocity and pressure unknowns $\bm{d}^n_b$ and $q^n_a$. Along with the basis functions, these yield the approximate velocity and pressure fields according to \eqref{eq:ansatzes}.

\subsection{Remarks}
\label{ssec:igacomments}

\begin{itemize}
\item Assuming incompressible material behavior, we have introduced a mixed discretization scheme. It is well known that the discretization spaces for the velocity and pressure fields must be chosen appropriately to ensure stability. In this paper, we adopt the subdivision approach proposed in \cite{RUBERG2012266}, where the B-Spline bases for the two fields are constructed in an efficient manner from equal polynomial order but different element subdivision, namely, in 2D one pressure element in the parametric space is split into $2\times 2$ velocity elements.

\item Integrals of a generic function $f$ over the physical domain at time step $t^n$ can be numerically approximated as
\begin{equation}
\int_{\Omega^n}f\left(\bm{x}\right)\mathrm{d}\Omega^n\approx\sum_{q=1}^Q W^n_q\cdot f\left(\bm{x}^n_q\right)
\label{eq:numericalintegration}
\end{equation}
In IGA, the $Q$ quadrature points $\bm{x}^n_q$ and quadrature weights $W^n_q$ are often chosen as Gauss quadrature, although much more efficient alternatives are available \citep{HUGHES2010301,FAHRENDORF2018390}.
 
\item In IGA, incorporation of (possibly inhomogeneous) Dirichlet boundary conditions in the discrete global equation system is a straightforward task due to the weak Kronecker-delta property of the basis functions, and is performed by imposing suitable control point values on boundary control points. More details on boundary conditions in IGA are given in \cite{COTTRELL200969}.

\item For the evolution of the physical domain in time, the control point velocities are integrated to displacements and added to previous control point positions, e.g. using a forward Euler scheme
\begin{equation}
\bm{c}^{n+1}_{ij} = \bm{c}^n_{ij}+\Delta \bm{c}^n_{ij} = \bm{c}^n_{ij}+ \bm{d}^n_{ij} \cdot\Delta t
\label{eq:igaeuler}
\end{equation}
Computing the basis functions at the next time step from the same bivariate B-Spline basis (which stays identical in the parametric space) but using updated control point positions attributes a material point character to all parametric points, i.e. a point in the parametric space is always associated to the same physical particle of moving material, which is favorable for the formulation of material time derivatives as well as for accurate integration. On the other hand, in presence of very large deformations, a severe distortion of the mapping between parametric and physical space occurs, and IGA basis functions in the physical domain may no longer be suitable for the analysis. This is qualitatively demonstrated in the upper branch of \autoref{fig:fligadistortion}. In such cases, techniques to control distortion of the basis functions become necessary.

\end{itemize}

\section{Basics of FLIGA}
\label{sec:FLIGA}

In this section, we introduce FLIGA as a new technique for controlling distortion of isogeometric B-Spline basis functions. FLIGA allows for the extension of isogeometric concepts to extreme deformation analysis. The previously reviewed fundamentals of classical Lagrangian IGA are all maintained, except for those concerning B-Spline basis function design and numerical Gauss quadrature. These are both modified in this section.

\subsection{Introduction of floating B-Splines}
\label{ssec:fligaintroduction}
\begin{figure}[h]
\centering
\includegraphics[trim={6cm 2cm 3.5cm 0.75cm},clip ,width=13.5cm]{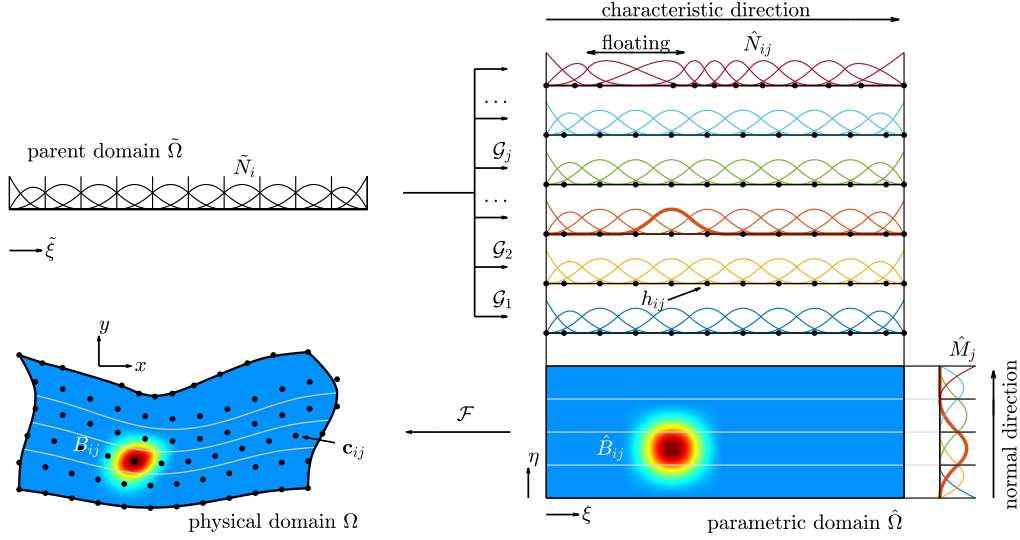}
\caption{Schematic overview of bivariate B-Spline basis construction in FLIGA.}
\label{fig:fligaconcept}
\end{figure}

We start the introduction of FLIGA by proposing an alternative for the construction of 2D B-Spline bases, as visualized in \autoref{fig:fligaconcept}.
The core idea is a generalization of the classical tensor-product structure from IGA. We denote this generalization as \textit{floating tensor-product structure}. On the parametric domain, we associate  each 1D B-Spline basis function $\hat{M}_{j}\in\left\lbrace\hat{M}_{j}\left(\eta\right)\right\rbrace^{J}_{j=1}$ with its \textit{own} 1D B-Spline basis $\left\lbrace\hat{N}^n_{ij}\left(\xi\right)\right\rbrace^{I}_{i=1}$, simply adding index $j$ to the latter.
\begin{equation}
\hat{B}^n_{ij}\left(\xi,\eta\right)=\hat{N}^n_{ij}\left(\xi\right)\hat{M}_{j}\left(\eta\right)
\label{eq:floatingtensorproduct}
\end{equation}
%For simplicity, we do not consider the case for which also the range $I$ depends on $j$, even though it might allow for interesting local refinement strategies.
In this floating tensor-product structure, $\xi$ is the \textit{characteristic} direction and $\eta$ is the \textit{normal} direction. In the physical space, the characteristic direction is mapped to the (possibly curved) direction along which deformations are expected to be severe, whereas the deformations along the normal direction are assumed to be moderate.  We denote as $\left\lbrace\hat{N}^n_{ij}\left(\xi\right)\right\rbrace^{I}_{i=1}$ the characteristic B-Spline bases, of which we count a total number of $J$. Similarly, the single basis $\left\lbrace\hat{M}_{j}\left(\eta\right)\right\rbrace^{J}_{j=1}$ is termed normal B-Spline basis. Note that the characteristic B-Spline bases in parametric coordinates are no longer the same throughout the analysis as in IGA, but evolve with proceeding time step $n$. This attribute extends to the 2D B-Spline basis $\left\lbrace\hat{B}^n_{ij}\left(\bm{\xi}\right)\right\rbrace^{I,J}_{i=1,j=1}$.

We derive the characteristic B-Spline bases from a static (i.e. unchanged throughout the analysis) 1D  \textit{parent} B-Spline basis $\left\lbrace\tilde{N}_{i}\left(\tilde{\xi}\right)\right\rbrace^{I}_{i=1}$ of order $r$ on $\tilde{\Omega}=\left[\tilde{\xi}_1,\tilde{\xi}_{I+r+1}\right]$ built by the recursive formula \eqref{eq:coxdeboor} with parent knot vector
\begin{equation}
\tilde{\Xi}=\left\lbrace \tilde{\xi}_{1},\tilde{\xi}_{2},\ldots,\tilde{\xi}_{i},\ldots,\tilde{\xi}_{I+r+1}\ \vert\ \tilde{\xi}_i \in \mathbb{R}\right\rbrace
\end{equation}
%Note that it does not depict the parent domain introduced in IGA literature for Gauss quadrature, though the idea to describe some reference pattern we derive individuals from is equivalent. The latter association is on purpose, and with quadrature being different in FLIGA, the term ``parent'' is free for our alternative use.
The term  \textit{parent} refers to the descendance of all characteristic bases from this basis. With the parent basis at hand, we introduce a set of $J$ linear mappings $\mathcal{G}^n_j: \tilde{\Omega}\rightarrow\hat{\Omega}^n_{\xi}$, with $\hat{\Omega}^n_{\xi}=[a^n,b^n]$, such that
\begin{equation}
\xi = \mathcal{G}^n_j\left(\tilde{\xi}\right)=\sum_{i} h^n_{ij}\tilde{N}_{i}\left(\tilde{\xi}\right)
\label{eq:Gmapping}
\end{equation}
We introduce the term \textit{floating maps} for these $\mathcal{G}^n_j$ in order to distinguish them from the later introduced geometric mapping to physical space. It is these floating maps that grant a floating character to the characteristic basis functions by translating the parent basis into the characteristic basis functions
\begin{equation}
\hat{N}^n_{ij}\left(\xi\right)=\tilde{N}_{i}\left(\mathcal{G}^{n^{-1}}_j\left(\xi\right)\right)
\label{eq:CharacteristicBases}
\end{equation}
As the arrangement of the linear combination coefficients $h^n_{ij}$ allows to control this floating, we refer to these as \textit{floating regulation points}. Due to weak Kronecker-delta property of the parent basis at its boundaries, the span of $\hat{\Omega}^n_{\xi}$ requires $\forall j: h^n_{1j}=a^n$ and $h^n_{Ij}=b^n$.
The scalar Jacobians of the floating maps are important quantities and read
\begin{equation}
J^n_j\left(\tilde{\xi}\right)=\sum_{i}h^n_{ij} \nabla_{\tilde{\xi}} \tilde{N}_{i}\left(\tilde{\xi}\right)
\label{eq:Gjacobian}
\end{equation}
Finally, the characteristic bases are combined by means of \eqref{eq:floatingtensorproduct} with the normal basis, see also \autoref{fig:fligaconcept}.

The procedure described above yields the 2D B-Spline basis supported on $\hat{\Omega}^n=\hat{\Omega}^n_{\xi}\times\hat{\Omega}_{\eta}$. Then, as in classical IGA, we use a geometric mapping $\mathcal{F}^n: \hat{\Omega}^n \rightarrow \Omega^n$ according to \eqref{eq:Fmapping} in order to obtain bivariate B-Spline bases defined in the physical space, see \eqref{eq:PhysicalB}.

The above procedure leads to the capability of basis functions corresponding to different normal parametric coordinates to evolve independently from each other along the characteristic direction. This feature underlies the terminology of \textit{floating} isogeometric analysis and proves crucial for the treatment of extreme deformations associated to the characteristic direction. Through the introduction of the floating regulation points, the degrees of freedom for basis function construction are significantly increased, however, as will be shown shortly, at a negligible computational cost. Let us assume having a parent and normal knot vector at hand. From the enriched basis function space that we obtain by introducing the additional mapping \eqref{eq:Gmapping}, we choose for the analysis the basis functions that are ideally suited in terms of distortion, while efficient to obtain. This leads to the updating schemes for the floating regulation points and the control points, which are given in Section \ref{ssec:updates}.

Unlike in standard IGA, in FLIGA we do not let the basis functions in physical space follow the deformation fully, but only as far as a sufficiently undistorted shape is preserved. Unlike in remeshing techniques, the evolution of the basis functions in FLIGA is incremental throughout the simulation. Conceptually, we lie in between and the recognition of the material deformation from basis function deformation is possible by averaging the movement of all basis functions. 

The FLIGA strategy we have described is reminiscent of core strategies in meshless methods, which also incrementally evolve basis functions while relaxing constraints from a rigid and Lagrangian element division. In FLIGA we loosen mesh division in the characteristic direction, however we preserve a generalization of the classical mesh, where elemental subdivision persists at the level of single basis functions. Thus, FLIGA can be viewed as a generalization of IGA to the floating tensor-product structure that we obtain by incorporating some aspects of meshless behavior.

\subsection{Floating regulation point updates}
\label{ssec:updates}

The proper choice of update rules for the position of control points and floating regulation points is essential for the success of FLIGA, as this combination offers control over the distortion of the basis functions in the physical space along the characteristic direction. By keeping for the control points,  $\bm{c}^n_{ij}$, the deformation-dependent update \eqref{eq:igaeuler} from classical IGA, the task reduces to finding an updating scheme for the floating regulation points, $h^n_{ij}$, such that distortion of basis functions is minimized. The update can be expressed as
\begin{equation}
h^{n+1}_{ij} = h^n_{ij} + \Delta h^n_{ij}
\end{equation}
and we need to determine $\Delta h^n_{ij}$. Let us introduce a time-invariant level function $\mathcal{L}$ which associates a desired parametric coordinate in the characteristic direction $\xi$ to each generalized physical coordinate $\bm{\chi}\in S^n\left(\bm{c}^n_{ij}\right)$, where $S$ denotes the polygonal area spanned by the control net $\left\lbrace\bm{c}^n_{ij}\right\rbrace^{I,J}_{i=1,j=1}$
\begin{equation}
\xi=\mathcal{L}\left(\bm{\chi}\right)
\end{equation}
Note that $\mathcal{L}$ cannot be chosen as the $\xi$-component of the inverse mapping $\mathcal{F}^{n^{-1}}$, since it has to be defined also for $\bm{\chi}=\bm{c}^n_{ij}\notin\Omega^n$. In the choice of $\mathcal{L}$ we have some freedom, provided it is sufficient regular in space and it has no local minima or maxima on all $S^n$.

Due to the association of floating regulation points and control points, an ideal update is
\begin{equation}
\mathrm{d}\xi=\nabla_{\bm{\chi}}\mathcal{L}(\bm{\chi})\cdot \mathrm{d}\bm{\chi}\ \ \ \ \ \ \ \  \rightarrow\ \ \ \ \ \ \ \  \Delta \bar{h}^n_{ij}=\nabla_{\bm{\chi}}\mathcal{L}(\bm{c}^n_{ij})\cdot \Delta\bm{c}^n_{ij}=\nabla_{\bm{\chi}}\mathcal{L}(\bm{c}^n_{ij})\cdot \bm{d}^n_{ij}\cdot \Delta t
\label{eq:floatregulationidealupdate}
\end{equation}
However, this in general would not preserve the rectangular structure of the parametric domain, as for $j=1,...,J$ different updates would be introduced on $h^n_{1j}$ (and also on $h^n_{Ij}$). To correct this, we introduce a blending function $\gamma^n\left( \xi \right)$ taking the value $\gamma^n=0$ at the boundaries $\xi=a^n$ and $\xi=b^n$, where we do not apply the update \eqref{eq:floatregulationidealupdate}, but one that preserves the rectangular geometry of $\hat{\Omega}$. A transition is defined towards the region $g^n_1<\xi<g^n_2$ in which a plateau is reached, $\gamma^n=1$. We typically choose $a^n<g^n_1\ll b^n$ and $a^n\ll g^n_2<b^n$ to obtain a reasonably large plateau in which updates are then fully addressed by \eqref{eq:floatregulationidealupdate}. For $0<\gamma^n<1$ a transitional treatment is applied.

Let us now apply $\mathcal{L}$ and $\gamma^n$ in order to obtain the incremental update of a floating regulation point $h^n_{ij}$ corresponding to the physical control point $\bm{c}^n_{ij}$ as
\begin{equation}
\Delta h^n_{ij}=\gamma^n\left(h^n_{ij}\right)\cdot \Delta \bar{h}^n_{ij} + \left(1-\gamma^n\left(h^n_{ij}\right)\right)\cdot \Delta \bar{h}^n_{(I^*)(J/2)}
\label{eq:floatregulationupdate}
\end{equation}
The uniform update at the boundaries $\xi=a^n$ and $\xi=b^n$ is thus achieved by employing for all $j=1,...,J$ the update \eqref{eq:floatregulationidealupdate} for the respective central boundary control point ($j=J/2$, rounded if required). To distinguish the boundaries $\xi=a^n$ and $\xi=b^n$, we choose in \eqref{eq:floatregulationupdate} $I^*=1$ for $i<I/2$ and $I^*=I$ for $i>I/2$. Note that the transition of the blending function between boundary and central regions in the characteristic direction ensures a proper distribution of the floating regulation points.

In the initial state, the floating regulation points $h^0_{ij}$ are chosen at the Greville abscissae \citep{AURICCHIO2010} of the parent knot vector, such that $\forall j: \ \xi=\mathcal{G}^0_j\left(\tilde{\xi}\right)=\tilde{\xi}$. As a result, the floating mesh of the initial time step $n=0$ conforms to a B-Spline mesh with classical element division. This ensures the compatibility of FLIGA with the isogeometric concept of using CAD geometry representations further for analysis.

\begin{figure}[h]
\centering
\includegraphics[width=0.4\textwidth]{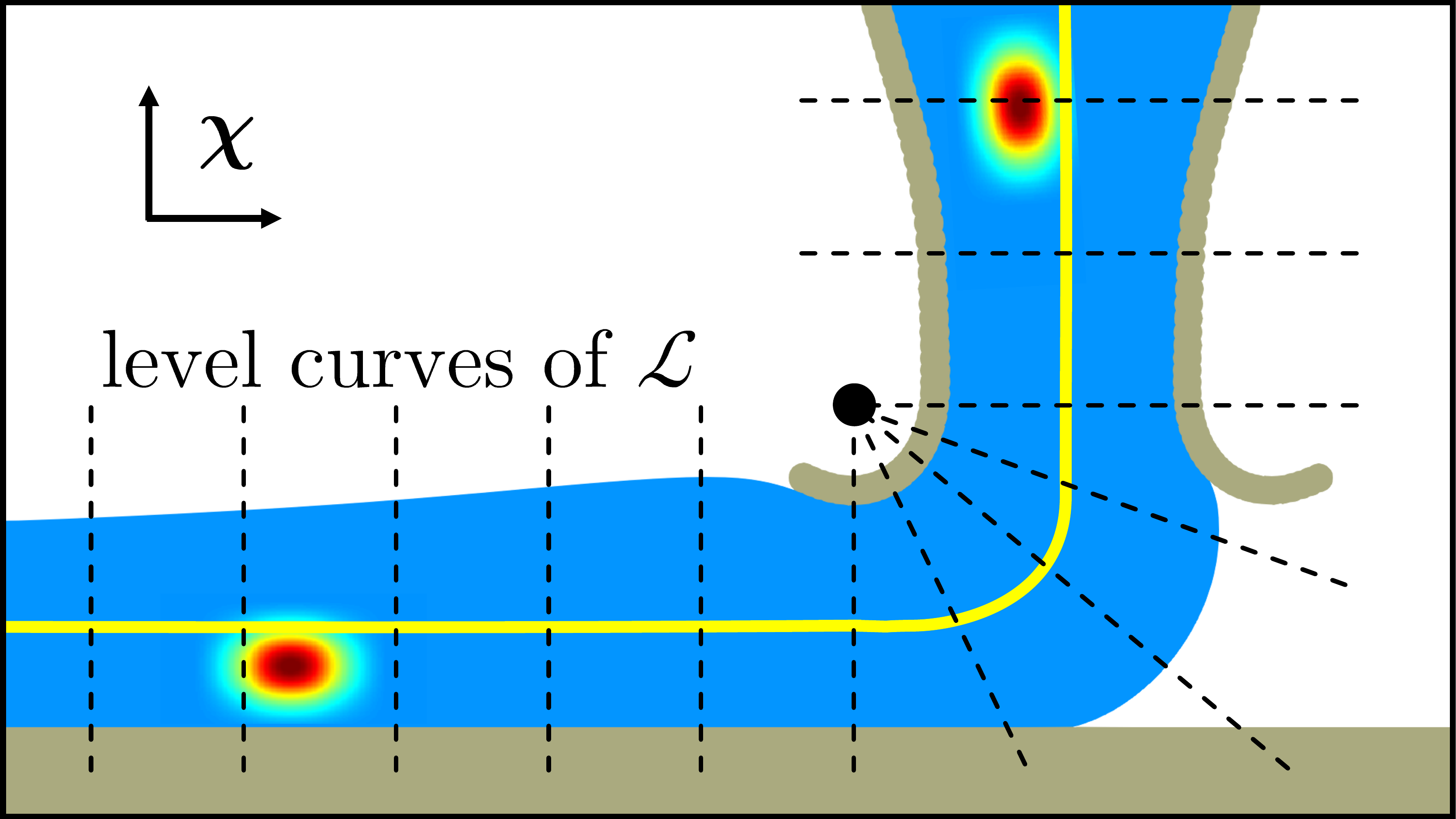}
\caption{Levels defined by a suitable level function for extrusion simulation.}
\label{fig:AM_Levelcurves}
\end{figure}

We illustrate the idea of the level function on the example of extrusion in \autoref{fig:AM_Levelcurves}, where the physical domain $\Omega^n$ is colored in blue. Obviously, a suitable mapping of the parametric space to $\Omega^n$ would associate the dashed lines in the figure to a constant $\xi$ each, with $\xi$ increasing from one dashed line to the next. Accordingly, we design $\mathcal{L}$ to take constant values along these levels and evolve monotonically along the characteristic extrusion direction (yellow). An update of the floating regulation points based on such $\mathcal{L}$ preserves the constancy of $\xi$ along the desired levels throughout simulation. We also depict the resulting behavior for two basis function candidates aligned to both the characteristic direction and the levels of $\mathcal{L}$. Note that the lack of regularity of the level function at the point evidenced with the black dot is unproblematic, as this point is never inside $\Omega^n$. Nonetheless, a certain distance between this or similar points and the physical domain is required.

\subsection{Basis function evaluation and connectivities}
\label{ssec:evaluation}

Let us now consider the location of a parametric point at time step $t^n$, $\bm{\xi}^n_q=\left(\xi^n_q,\eta^n_q\right)^T$, and aim at computing the basis functions, their physical derivatives and the control point connectivity. 

We begin by identifying the normal connectivities $\mathcal{J}^n_q=\left\lbrace j \in \left\lbrace 1,...,J\right\rbrace : \hat{M}_j(\eta^n_q)\neq0 \right\rbrace$ associated to $\bm{\xi}^n_q$. Due to the classical 1D B-Spline basis character of the normal basis, this task can be performed just like in IGA. As the considered parametric point might in general pass element boundaries in normal direction during the simulation, these connectivities must be obtained for the current time step.

Then, for each $j \in \mathcal{J}^n_q$, the corresponding parent coordinate
\begin{equation}
\tilde{\xi}^n_{qj}=\mathcal{G}^{n^{-1}}_j\left(\xi^n_q\right)
\label{eq:inverseG}
\end{equation}
is computed by means of Newton-Raphson schemes, see \ref{parentnewton} for details.

Given the different parent coordinates $\tilde{\xi}^n_{qj}$ and the unique normal coordinate $\eta^n_q$, we can compute the parametric basis function values at $\bm{\xi}^n_q$ as
\begin{equation}
\hat{B}^n_{ij}(\bm{\xi}^n_q)=\hat{N}^n_{ij}(\xi^n_q)\cdot\hat{M}_j(\eta^n_q)=\tilde{N}_i(\tilde{\xi}^n_{qj})\cdot\hat{M}_j(\eta^n_q)
\label{eq:parametricbsplinesevaluated}
\end{equation}
while the scalar Jacobians \eqref{eq:Gjacobian} yield the parametric derivatives
\begin{equation}
\begin{split}
\nabla_{\bm{\xi}}\hat{B}^n_{ij}(\bm{\xi}_q^n)&=\begin{pmatrix}
\nabla_{\xi}\hat{N}^n_{ij}(\xi^n_q)\cdot\hat{M}_j(\eta^n_q)\\
\hat{N}^n_{ij}(\xi^n_q)\cdot\nabla_{\eta}\hat{M}_j(\eta^n_q)\\
\end{pmatrix}\\&
=\begin{pmatrix}
J^n_j\left(\tilde{\xi}^n_{qj}\right)^{-1}\cdot\nabla_{\tilde{\xi}} \tilde{N}_{i}\left(\tilde{\xi}^n_{qj}\right)\cdot\hat{M}_j(\eta^n_q)\\
\tilde{N}_i(\tilde{\xi}^n_{qj})\cdot\nabla_{\eta} \hat{M}_{j}\left(\eta^n_q\right)\\
\end{pmatrix}
\end{split}
\end{equation}
The mapping of the evaluated B-Spline basis functions to the physical point $\bm{x}^n_q=\mathcal{F}^n\left(\bm{\xi}^n_q\right)$ does not alter the basis function value
\begin{equation}
B^n_{ij}(\bm{x}^n_q)=\hat{B}^n_{ij}(\bm{\xi}^n_q)
\label{eq:physicalbsplinesevaluated}
\end{equation}
but controls the physical gradients
\begin{equation}
\nabla_{\bm{x}}B^n_{ij}\left(\bm{x}^n_q\right)=\bm{J}^n\left(\bm{\xi}^n_q\right)^{-T}\nabla_{\bm{\xi}}\hat{B}^n_{ij}(\bm{\xi}^n_q)
\end{equation}
via the Jacobian \eqref{eq:Fjacobian}.

For assembling the contribution of the evaluation point into the global system of equations, its global control point connectivities are required. We complement the normal connectivities $\mathcal{J}^n_q$ by the $J$ sets of characteristic connectivities $\mathcal{I}^n_{qj}=\left\lbrace i \in \left\lbrace 1,...,I\right\rbrace: \tilde{N}_i(\tilde{\xi}^n_{qj})\neq 0\right\rbrace$, obtained again as in IGA. The global connectivities of the FLIGA B-Spline basis are then defined as
\begin{equation}
\mathcal{B}^n_q=\left\lbrace b \in \left\lbrace 1,...,I\cdot J\right\rbrace : \hat{B}_{ij}(\bm{\xi}^n_q)\neq0 \right\rbrace=\left\lbrace b=(j-1)\cdot I+i\ \left.\right|\ i \in \mathcal{I}^n_{qj},j \in \mathcal{J}^n_q \right\rbrace.
\end{equation}
Profiting from the parametric substructure, this control point search procedure is significantly simpler than nodal search algorithms of classical meshless methods.

\subsection{Basis function properties}

As follows, we list and discuss the most important properties of FLIGA B-Splines in order to assess their suitability for analysis:

\begin{itemize}
\item  \textit{Continuity:} Basis function continuity follows the same rules as for classical tensor-product B-Splines. However, in FLIGA, the lower-continuity locations of the univariate B-Splines are not all aligned in the normal parametric direction as in IGA, see \ref{continuityandsupport} for an exemplary illustration.
\item \textit{Support:} The number of supported basis functions at an arbitrary evaluation point is constant over the entire domain and is the same as in IGA - e.g. for splines of order $r$ in both parametric directions, this number is $(r+1)^2$.
\item \textit{Partition of unity:} The proof can be found in \ref{partitionofunity}.
\item \textit{First-order consistency:} It is demonstrated with a simple patch test in Section \ref{ssec:patchtest}.
\item \textit{Weak Kronecker-delta property:} The proof can be found in \ref{weakkroneckerdeltaproperty}.
\item \textit{Boundary preservation:} Physical boundaries are preserved while basis functions are floating; see \ref{boundary preservation} for a proof.
\end{itemize}

The above list shows that FLIGA B-Splines preserve many attributes of classical tensor-product B-Splines that are favorable for analysis. For instance, a minimum order of global continuity can be prescribed by the choice of the polynomial B-Spline degree. Continuity can thus be adapted to the problem at hand. The fact that the number of supported basis functions at an evaluation point is constant over the entire domain and known \textit{a priori} is convenient for the implementation, as there is no need to adaptively tune parameters to adjust the overlap of the basis functions during the simulation (as sometimes necessary in meshless methods). In FLIGA, such overlap is straightforward to identify on the parametric domain, hence the search for connectivities can be carried out with lower effort than in meshless methods. Mixed discretization pairs can be easily obtained by adjusting the polynomial orders for velocity and pressure (e.g. as Taylor-Hood pair) or by properly subdividing parent and normal domain (as mentioned earlier, we apply the latter strategy in this work, see Section \ref{ssec:mixedsubdivisionFLIGA}). FLIGA B-Splines always exactly cover the analysis domain, allowing for the natural treatment of moving boundaries as in Lagrangian IGA. Boundaries are precisely defined in FLIGA and can be consistently refined with strategies from IGA. The imposition of essential boundary conditions is straightforward due to the weak Kronecker-delta property. First-order consistency is naturally satisfied, whereas it is not present in many meshless techniques. In fact, the lack of (weak) Kronecker-delta property and first-order consistency is believed to be related to most numerical issues encountered in the meshless context \citep{Fougeron2018}.

However, FLIGA also has limitations, most importantly, the restriction that mesh distortion can only be overcome in one parametric direction. The fact that the parametric domain must have rectangular geometry is a limitation inherited from IGA that can be partially solved by the extension to multi-patch geometries - however the freedom of meshless methods in dealing with complex topologies and/or topological changes cannot be achieved. Also, refinement is affected by the anisotropic nature of basis function construction and thus is non-local in the normal direction. Finally, exact quadrature of FLIGA B-Splines is also challenging, see Section \ref{ssec:materialpointintegration}.

\subsection{Mixed subdivision technique}
\label{ssec:mixedsubdivisionFLIGA}
Since our intention is to apply FLIGA to the study of incompressible materials (the polymer melt in extrusion-based AM), we need mixed-field approximations to avoid volumetric locking. Thus, we employ the previously introduced design principles to obtain a pair of FLIGA B-Spline bases: one for geometry and velocity (as per the isoparametric concept), and the other one for pressure approximation. 

\begin{figure}[h]
\centering
\includegraphics[trim={7cm 2cm 5cm 0.75cm},clip ,width=13.5cm]{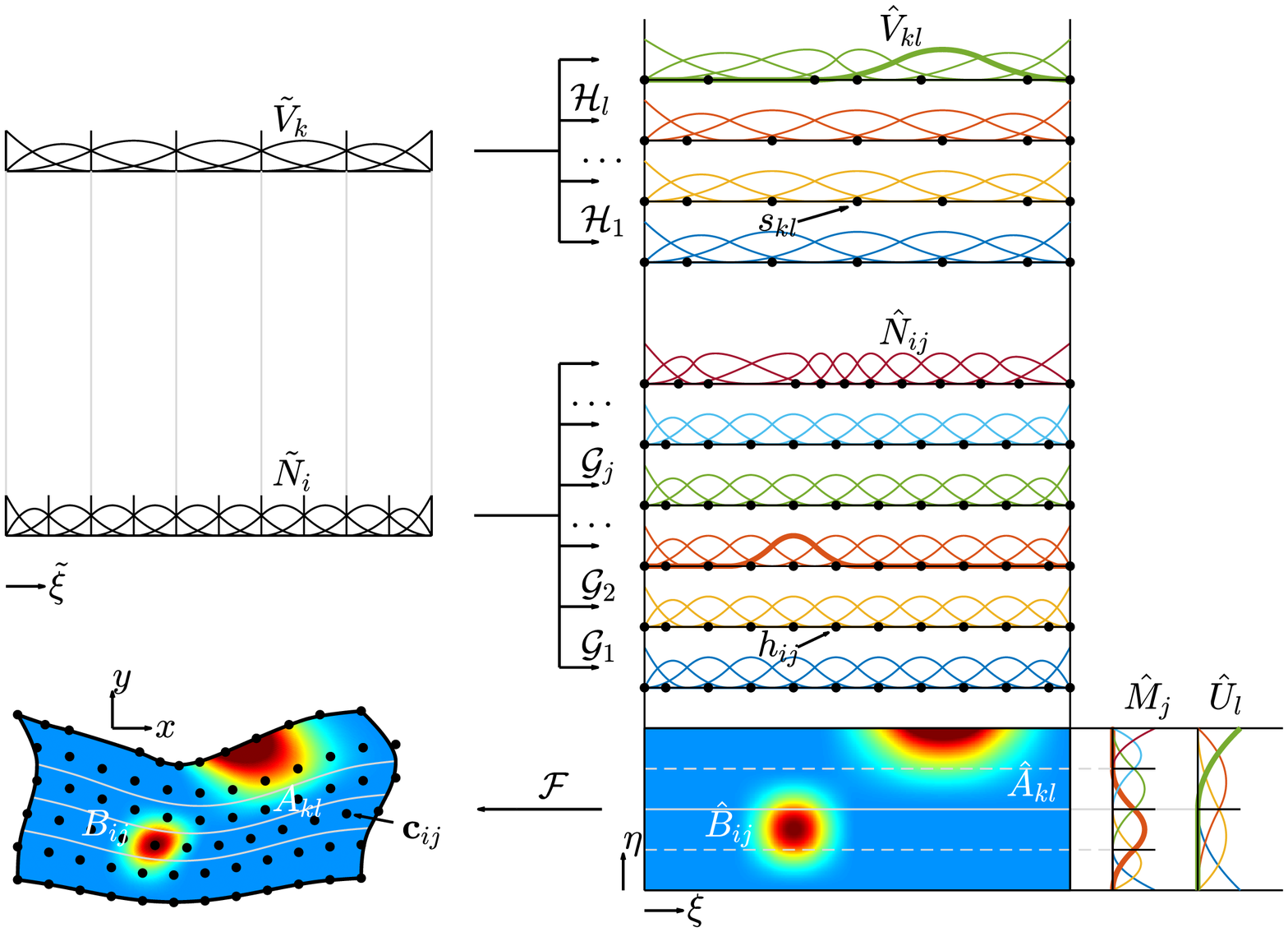}
\caption{Schematic overview of the mixed subdivision technique in FLIGA}
\label{fig:fligaconceptSUB}
\end{figure}

For the choice of the velocity and pressure bases, we propose an adaption of the mixed subdivision strategy from classical IGA \citep{RUBERG2012266} to FLIGA, see the illustration in \autoref{fig:fligaconceptSUB}. We construct the 2D pressure approximation from coarser parent and normal bases, $\left\lbrace\tilde{V}_k\right\rbrace^K_{k=1}$ and $\left\lbrace\hat{U}_k\right\rbrace^L_{l=1}$, respectively, than the equivalents for velocity, $\left\lbrace\tilde{N}_i\right\rbrace^I_{i=1}$ and $\left\lbrace\hat{M}_j\right\rbrace^J_{j=1}$. In our computations, locking is successfully suppressed if one pressure parent knot span comprises two equal-sized parent knot spans for velocity and the same holds for the normal bases. Based on these observations, we propose  the use of an equal parent domain $\tilde{\Omega}$ for velocity and pressure.

Then, we introduce one set of floating regulation points $h^n_{ij}$ for design of the velocity basis, as well as another set $s^n_{kl}$ for design of the pressure basis. The corresponding floating maps $\mathcal{G}^n_j$ and $\mathcal{H}^n_l$ are unique to the velocity or pressure basis, respectively.

Unlike the floating maps, the physical mapping is universal and determined by the velocity basis and the physical control points $\bm{c}^n_{ij}$. Thus, the velocity and pressure bases must be constructed on the same parametric domain $\hat{\Omega}^n$. Note that, despite the coarser division, the physical boundaries of the pressure basis coincide with those of the velocity basis.

We can now apply the update procedure of Section \ref{ssec:updates} to the control points $\bm{c}^n_{ij}$ and velocity floating regulation points $h^n_{ij}$ and introduce another update strategy for aligning the choice of $s^n_{kl}$ in a subsequent step by enforcing $\forall l$:
\begin{equation}
\xi_{ml}=\sum_{k} \tilde{V}_{k}\left(\tilde{\xi}_{m}\right) \cdot s^n_{kl}=\sum_{i} \tilde{N}_{i}\left(\tilde{\xi}_{m}\right) \cdot h^n_{ij}=\xi_{mj}  \ , \ \ \ \ \ \ \ \ j=1+\mathrm{round}\left(\dfrac{l-1}{L-1}(J-1)\right)
\label{eq:floatregulationpressure}
\end{equation}

The $\tilde{\xi}_{m}$ with $m=1,...,K$ can be considered as collocation points in the parent domain, which we choose as Greville abscissae. In this way, we determine all $s^n_{kl}$ such that the parent collocation points are mapped to the same parametric characteristic points $\xi_{ml}$ as the floating maps for velocity would do. To this end, due to the subdivision of the normal basis functions, the indices $l$ and $j$ must be associated, here for simplicity by an average scaling on the index numbers. The $s^0_{kl}$ in the initial step may already be derived by the collocation scheme given above and satisfy similarly to the velocity floating maps $\forall l: \ \xi=\mathcal{H}^0_l\left(\tilde{\xi}\right)=\tilde{\xi}$.

\subsection{Material point integration}
\label{ssec:materialpointintegration}
As mentioned earlier, a parametric point (e.g. a Gauss point) in IGA is always associated to the same material point during its movement through time and physical space. In FLIGA, one natural consequence of breaking the Lagrangian character of the physical mapping is that one physical material point occupies different positions in the parametric and in the parent space during the course of the analysis. Thus, we cannot preserve both material and Gauss point character at the same time.

In FLIGA, we employ the material point integration strategy, which is well known from meshless methods, see, e.g. \cite{KUMAR2019858,KUMAR2020109329}. This technique attributes to the quadrature points a Lagrangian particle character in the sense of \eqref{eq:lagrangianviewpoint}. In the following, we elaborate on the derivation of these points and their quadrature weights, defining integration by \eqref{eq:numericalintegration}.

At the initial time step $n=0$, we construct a standard Gauss integration scheme for the velocity mesh, which corresponds to a classical IGA mesh at this stage. This provides parametric material points $\bm{\xi}^0_q$ and parametric weights $\hat{W}^0_q$, which are well suited for quadrature at this time step. We map these to the physical domain, obtaining
\begin{equation}
\begin{split}
\bm{x}^0_q &= \mathcal{F}^0\left( \bm{\xi}^0_q\right)\\
W^0_q &= \hat{W}^0_q \cdot \mathrm{det} \left(\bm{J}^0 \left(\bm{\xi}^0_q\right)\right)
\end{split}
\end{equation}
We then evolve this quadrature set in the physical space during the analysis, following the deformation of the material
\begin{equation}
\begin{split}
\bm{x}^{n+1}_q &= \bm{x}^n_q+ \bm{v}^n\left(\bm{x}^n_q\right)\cdot \Delta t\\
W^{n+1}_q &= W^n_q \cdot \mathrm{det} \left(\bm{\mathrm{I}}+ \nabla_{\bm{x}}\bm{v}^n\left(\bm{x}^n_q\right)\cdot \Delta t\right)
\end{split}
\label{eq:mpupdate}
\end{equation}
In order to obtain the parametric material point positions at which the basis functions have to be evaluated according to Section \ref{ssec:evaluation}, we have to map the physical positions back to the parametric space
\begin{equation}
\bm{\xi}^n_q = \mathcal{F}^{n^{-1}}\left(\bm{x}^n_q\right)
\label{eq:mapbackmp}
\end{equation}
This is accomplished by a Newton-Raphson scheme described in \ref{materialpointnewtonraphson}.

Note that material point integration may result in excessive and/or highly irregular spacings between material points when deformations are too complex. Such phenomena also affect material point methods \citep{YUE2015}, and remedies from such methods may be adapted to FLIGA in future work. In any case, for the purpose of the present investigation even standard material point integration appears well suited. This is because, when a characteristic direction dominates the deformation as assumed in this work, the particle distribution typically stays relatively regular. Finally, material point integration in FLIGA clearly profits from the similarly deformation-dependent update of control points, which ensures a compatible alignment of material points with control points throughout the simulation.

The absence of advective terms makes material point integration particularly attractive for viscoelastic, inertial or non-isothermal flow, which otherwise would often require stabilization. Material derivatives from the continuous model are naturally discretized by
\begin{equation}
\left.\dfrac{D}{Dt}(\bullet)_q\right|_{t=t^{n-1}}\approx\dfrac{(\bullet)^n_q-(\bullet)^{n-1}_q}{\Delta t}
\label{eq:matdermp}
\end{equation}
where $(\bullet)_q$ and $(\bullet)^n_q$ are the respective continuous and discrete quantity carried by the material point $\bm{x}_q$.

\subsection{Code structure}
\label{ssec:proceduralstructure}

\normalem
\begin{algorithm}[t]
\DontPrintSemicolon
\KwIn{Problem data}
\SetKwBlock{Begin}{Procedure:}{end function}
\Begin(Floating isogeometric analysis)
{
define: knot vectors; Greville points $\tilde{\xi}_m$; polynomial order: $r$; initial configuration:  $h^0_{ij}$, $\bm{c}^0_{ij}$, $\bm{x}^0_q$, $W^0_q$\;
provide tools for velocity floating regulation point update: $\nabla_{\bm{\chi}}\mathcal{L}$, $\gamma^n$\;
	\For {$\mathrm{\ time\ step}\ n=0,1,\ldots$}
	{
		compute floating regulation points for pressure: $s^n_{kl}$\;
		compute parametric material points: $\bm{\xi}^n_q$\;
		compute basis functions and connectivities at each material point: $A^n_{a}$, $B^n_{b}$, $\nabla_{\bm{x}}A^n_{a}$, $\nabla_{\bm{x}}B^n_{b}$, $\mathcal{A}^n$, $\mathcal{B}^n$\;
		update definition of boundary conditions\;
		\For {$\mathrm{Newton}$-$\mathrm{Raphson\ step}\ r=0,1,\ldots$}
		{
			\For {$\mathrm{material\ point\ index}\ q=0,1,\ldots,Q$}
			{
				assemble material point contribution to residual vector: $\bm{R}^{n,r}\mathrel{+}=\bm{R}^{n,r}_q$\;
			}
			apply Dirichlet boundary conditions on $\bm{R}^{n,r}$\;
			\uIf {$\Vert \bm{R}^{n,r}\Vert < \mathrm{tol}$}
			{
				time step solution: $\bm{u}^n\leftarrow\bm{u}^{n,r}$\;
				leave Newton-Raphson loop\;
			}
			\For {$\mathrm{material\ point\ index}\ q=0,1,\ldots,Q$}
			{
				assemble material point contribution to tangent stiffness matrix: $\bm{K}^{n,r}\mathrel{+}=\bm{K}^{n,r}_q$\;
			}
			apply Dirichlet boundary conditions on $\bm{K}^{n,r}$\;
			update solution guess by solving Newton-Raphson step: $\bm{u}^{n,r+1}=\bm{u}^{n,r}+\left({\bm{K}^{n,r}}\right)^{-1}\bm{R}^{n,r}$\;
		}
		update: $\bm{c}^{n+1}_{ij}$, $h^{n+1}_{ij}$, $\bm{x}^{n+1}_q$, $W^{n+1}_q$, internal variables\;
	}
}
\KwOut{Approximate fields: $\bm{v}^n, p^n$}
\caption{Analysis procedure}\label{FLIGAalgorithm}
\end{algorithm}
\ULforem

We conclude the introduction of FLIGA by brief remarks on a potential code structure, see \autoref{FLIGAalgorithm}.

The initial step of the analysis procedure is the collection of relevant data, i.e. parameters specifying material properties, loading conditions and numerical parameters. Afterwards, parent and normal knot vectors are set up according to the desired element subdivision, and the Greville point set $\tilde{\xi}_m$ for the later updates \eqref{eq:floatregulationpressure} of the pressure floating regulation points is created. The definition of the polynomial order, of the initial floating regulation points for velocity as well as of the initial control points completes the geometry preparation for the first time step. Also, we distribute material points as Gaussian quadrature points of this initial classical IGA mesh. Lastly, we design a suitable level function and blending function, which will be required for the later updates \eqref{eq:floatregulationupdate} of the velocity floating regulation points. Note that the design of level and blending function is not automatic, but requires some knowledge on the expected deformations.

Having specified the problem setup, a time step loop is started. Floating regulations points for pressure are computed from those for velocity \eqref{eq:floatregulationpressure}. Parametric material points are obtained by mapping back current physical positions of the material points \eqref{eq:mapbackmp}. This provides the information to compute for each material point the values of the supported basis functions and the control point connectivity (Section \ref{ssec:evaluation}). The treatment of Dirichlet boundary conditions is identical as in IGA (Section \ref{ssec:igacomments}).

For solving the global system of discrete equations (Section \ref{ssec:discreteequations}), we apply a Newton-Raphson procedure (\ref{globalNewtonRaphson}) and therefore introduce a nested iteration loop. Within this loop, we first compute the residual vector for a current guess of the unknowns. Other than in IGA, we do not assemble quadrature point contributions to intermediate element vectors. Instead, assembly occurs directly into the global system. To this end, we do not iterate over elements, but once over all material points. Contributions are computed for the global system \eqref{eq:globalsystem1} and \eqref{eq:globalsystem2}, where we approximate integrals by \eqref{eq:numericalintegration} and apply material point integration (Section \ref{ssec:materialpointintegration}). Dirichlet boundary conditions are now incorporated to the global residual vector.

Next, the residual norm is computed. If this norm lies below a prescribed tolerance, the Newton-Raphson procedure is considered converged. Otherwise, we assemble all material point contributions to the tangent stiffness matrix in similar manner as previously done for the residual vector. Dirichlet boundary conditions are applied. Finally, the guess of the unknowns is improved and the next Newton-Raphson iteration initiated. This process repeats until the residual tolerance is met.

Upon convergence, the current time step is finalized by updating the locations of control points \eqref{eq:igaeuler}, velocity floating regulations points \eqref{eq:floatregulationupdate}, and material points, as well as the material point weights \eqref{eq:mpupdate} and possibly the internal variables from the constitutive laws (see Section \ref{ssec:constitutivelaws}). This step requires the evaluation of the level function and current blending function.

Finally, the next time step is initiated and the procedure is repeated.

\section{Numerical examples}
\label{sec:results}

In this section, we provide insight into the numerical accuracy and stability of FLIGA using three numerical examples: patch test, Taylor-Couette flow, and polymer extrusion. Preliminarily, we specify two constitutive models to be adopted in the examples.

\subsection{Constitutive models}
\label{ssec:constitutivelaws}
In the numerical examples, we will use two material models, namely, the Newtonian fluid and the viscoelastic Oldroyd-B fluid, which are briefly described as follows. For details on their theoretical background we refer to \cite{PHANTHIEN2013}.
In the Newtonian fluid model, $\bm{\tau}=\bm{\tau}_s$ is linearly related to the strain rate tensor
\begin{equation}
\bm{\tau}_s = 2 \eta_s \bm{D}
\label{eq:newtonianfluid}
\end{equation}
where $\eta_s$ denotes the Newtonian (or solvent) viscosity.
The viscoelastic Oldroyd-B model is more complex and accounts for viscous and elastic components. Even under large deformations, it appears well suited for representing material behavior in certain polymer deformation problems. In this model, $\bm{\tau}=\bm{\tau}_{OldB}$ is split into two contributions:
\begin{equation}
\bm{\tau}_{OldB} = \bm{\tau}_s+\bm{\pi}
\end{equation}
While $\bm{\tau}_s$ can immediately be adopted from the solvent Newtonian fluid model, $\bm{\pi}$ accounts for a polymeric contribution and is given in terms of the differential equation in time
\begin{equation}
\dfrac{D}{Dt}\bm{\pi}=2\eta_p\lambda^{-1}\bm{D}-\lambda^{-1}\bm{\pi}+ \nabla_{\bm{x}} \bm{v}\cdot\bm{\pi}+ \bm{\pi}\cdot\nabla_{\bm{x}}^T \bm{v}
\label{eq:oldroydevolution}
\end{equation}
Here, $\lambda$ is the relaxation modulus and $\eta_p$ is the polymeric viscosity. We require an initial condition $\bm{\pi}\left(t=0\right)=\bm{\pi}^0$ to complete the problem statement.

Let us now seek the discretized versions of the deviatoric stresses defined above. From now on, we will remain in the discretized setting. 
In the case of the Newtonian fluid, the task is trivial. Here suffices a replacement of the continuous strain rate in \eqref{eq:newtonianfluid} by the discretized strain rate $\bm{D}^n$. This yields the discretized solvent deviatoric stress $\bm{\tau}^n_s$.
For the Oldroyd-B model, we further require the discretized polymeric deviatoric stress $\bm{\pi}^n$. Using \eqref{eq:matdermp} for material point integration, we discretize \eqref{eq:oldroydevolution} obtaining $\bm{\pi}^n$ at material point $\bm{x}_q$ as
\begin{equation}
\bm{\pi}^n_q=\bm{\pi}^{n-1}_q+\dfrac{\Delta t}{\lambda} \left[ 2\eta_p\bm{D}^{n-1}_q-\bm{\pi}^{n-1}_q\right] + \Delta t \left[\nabla_{\bm{x}}\bm{v}^{n-1}_q\cdot\bm{\pi}^{n-1}_q+\bm{\pi}^{n-1}_q\cdot \nabla_{\bm{x}}^T\bm{v}^{n-1}_q \right]
\end{equation}

These computations do not require unknown information from the current time step, yielding an explicit time integration. History information from the previous step is available and stored as internal variable at the material point. The total deviatoric Oldroyd-B stress $\bm{\tau}^{n}_{OldB}$ is finally the sum of discrete solvent and discrete polymeric stress contribution.

For setting up the tangent stiffness matrix, the linearizations of the deviatoric Cauchy stresses $\bm{\tau}^n_s$ and $\bm{\tau}^n_{OldB}$ w.r.t. the current control point velocities are required, see \ref{deviatoriccauchylinearizations}. Assembling all information, we give a compact description of the global tangent stiffness matrix for implementation purposes in \ref{finalstiffnesstangent}. The governing equations are linear in control point velocity and pressure. However, the final system may still lose its linearity after application of nonlinear boundary conditions.

Alternative to the explicit time integration scheme proposed here, one could also apply an implicit analog. Being unconditionally stable, such an implicit formulation would lower the restrictions on the time step size. However, this introduces non-linearity in the governing equations and requires a more demanding linearization.

\subsection{Patch test}
\label{ssec:patchtest}

\begin{figure}[t]
\centering
\includegraphics[trim={5.25cm 2.9cm 2.5cm 2.25cm},clip,width=0.65\textwidth]{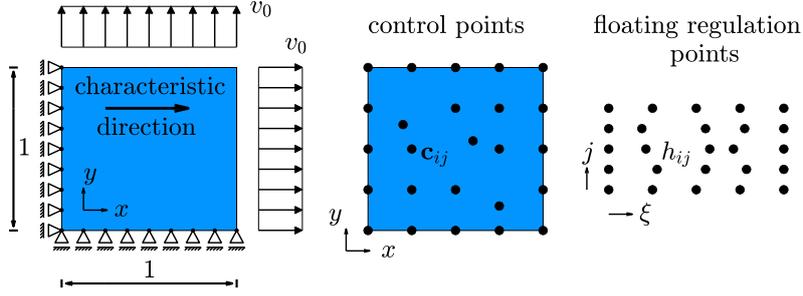}
\caption{Patch test problem for FLIGA.}
\label{fig:patchtest}
\end{figure}

We begin our numerical investigations by a patch test in order to assess first-order consistency. For simplicity, we adopt dimensionless units.
A square domain of Newtonian fluid ($\eta_s=50$) is subject to the boundary conditions illustrated in \autoref{fig:patchtest}, with $v_0=1$. We solve \eqref{eq:globalsystem1}, where we set the pressure to zero and thus do not need the incompressibility constraint \eqref{eq:globalsystem2}. The analytical velocity solutions are given by the monomials
\begin{equation}
\begin{pmatrix}
v_x\\
v_y
\end{pmatrix}
=
\begin{pmatrix}
x\\
y\\
\end{pmatrix}
\end{equation}

We construct FLIGA discretizations where the characteristic direction is associated to the $x$-axis. We distribute arbitrary $5\times 5$ floating regulation points on $\hat{\Omega}_{\xi}$ and associate an equivalent number of $5\times 5$ control points. The control points are likewise positioned arbitrarily, where their irregular distribution is the same for all studied polynomial orders. According to the constant control point number, we derive for each of these polynomial orders equal-sized element divisions for parent and normal domain. Quadrature is carried out by material point integration with either a moderate or an overkill number of points. It is not possible to define a classical Gauss point set for the floating meshes, however, we choose for the material points only those positions that are Gauss points of at least \textit{one} supported basis function. For comparison, we also test classical IGA with Gauss integration, for which the floating regulation points are all equidistantly aligned in order not to introduce a floating of basis functions.

Results of the patch test are given in \autoref{tab:patchtest} and show the logarithm of the $L^2$ norm of the relative error for the approximate velocity solutions. Classical IGA passes the test for all polynomial orders $r$, all quadrature sets and both velocity components. On the other hand, we find FLIGA to fulfill the patch test only when overkill integration is applied. This indicates that FLIGA B-Splines fulfil first-order consistency up to the error induced by the numerical quadrature.

For overkill integration of linear FLIGA B-Splines we do not provide specific values but the observation that the $L^2$ norm of the relative error converges towards zero with errors being $\mathcal{O}(Q^{-1})$ in the point number $Q$. For instance, $\left\lbrace 10^2,10^3,10^4,...\right\rbrace$ quadrature points yield logarithmic errors of approximately $\left\lbrace-2,-3,-4,...\right\rbrace$. We checked this behavior up to errors of $-8.5$, a further increase of quadrature points being exceedingly expensive.

\begin{table}[tb]
\centering
\begin{tabular}{l l l c l l l l l} 
 \toprule
 \multicolumn{2}{l}{Polynomial order} & \multicolumn{2}{c}{r=1} & \multicolumn{2}{c}{r=2} & \multicolumn{2}{c}{r=3} \\
  \multicolumn{2}{l}{No. of quadrature points} & $\approx 1\mathrm{e}2$ & $\rightarrow \infty$ & $\approx 1\mathrm{e}2$ & $\approx 1\mathrm{e}6$   & $\approx 1\mathrm{e}2$ & $\approx 1\mathrm{e}6$   \\ [0.5ex] 
 \hline
 \multirow{2}{4em}{IGA} & $x\ \mathrm{direction}$ & -16.28 &  & -15.67 & -14.80 & -15.73 & -14.47 \\
  & $y\ \mathrm{direction}$ & -16.02 &  & -15.85 & -13.69 & -15.63 & -14.26 \\
  \hline
  \multirow{2}{2em}{FLIGA} & $x\ \mathrm{direction\ (characteristic)}$ & -1.93 &  \multirow{2}{4em}{lin. con- vergence} & -2.46 & -10.76 & -2.88 & -11.73 \\
 & $y\ \mathrm{direction\ (normal)}$ & -2.86 & & -3.78 & -12.18 & -3.94 & -12.19 \\
 \bottomrule
\end{tabular}
\caption{Logarithmic relative $L^2$ errors of velocities in the patch test.}
\label{tab:patchtest}
\end{table}

%\begin{table}[tb]
%\centering
%\begin{tabular}{l l l l c l l l l l} 
% \toprule
% \multicolumn{3}{l}{polynomial order} & \multicolumn{2}{c}{r=1} & \multicolumn{2}{c}{r=2} & \multicolumn{2}{c}{r=3} \\
%  \multicolumn{3}{l}{quadrature points} & $\approx 1\mathrm{e}2$ & $\rightarrow \infty$ & $\approx 1\mathrm{e}2$ & $\approx 1\mathrm{e}6$   & $\approx 1\mathrm{e}2$ & $\approx 1\mathrm{e}6$   \\ [0.5ex] 
% \hline
% \multirow{2}{4em}{IGA} & \multirow{2}{4em}{irregular} & case I & -16.28 &  & -15.67 & -14.80 & -15.73 & -14.47 \\
%  & & case II & -16.02 &  & -15.85 & -13.69 & -15.63 & -14.26 \\
%  \hline
%  \multirow{4}{4em}{FLIGA} & \multirow{2}{4em}{reparam.} & case I & \multicolumn{1}{l}{-2.00} & \multirow{4}{4em}{lin. con- vergence} & -2.00 & -9.96 & -2.84 & -11.77 \\
% & & case II & \multicolumn{1}{l}{-2.43} &  & -3.77 & -11.53 & -3.33 & -12.16 \\
% & \multirow{2}{4em}{irregular} & case I & \multicolumn{1}{l}{-1.93} &  & -2.46 & -10.76 & -2.88 & -11.73 \\
% & & case II & \multicolumn{1}{l}{-2.86} & & -3.78 & -12.18 & -3.94 & -12.19 \\
% \bottomrule
%\end{tabular}
%\caption{Logarithmic relative $L^2$ errors for the traction test.}
%\label{tab:tractiontest}
%\end{table}

\subsection{Taylor-Couette flow: general remarks}

\begin{figure}[bt]
\centering
\includegraphics[width=0.4\textwidth]{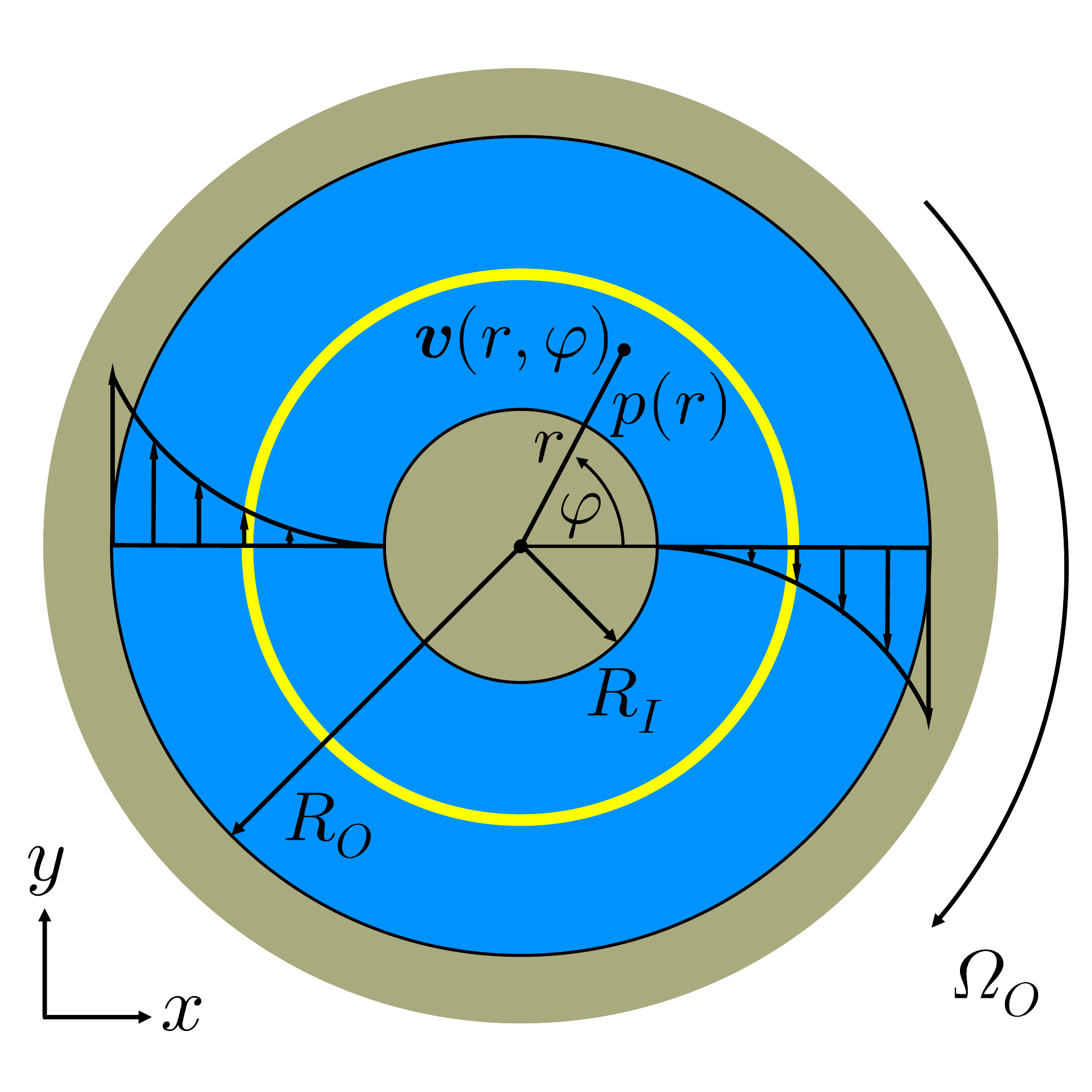}
\caption{Taylor-Couette problem.}
\label{fig:taylorcouette}
\end{figure}

We now consider the benchmark problem of inertialess Taylor-Couette flow, see the setup in \autoref{fig:taylorcouette}. We study the flow behavior of a fluid comprised between two concentric cylinders, with inner and outer radii $R_I$ and $R_O$, respectively, see \autoref{tab:TCparameters}. We let the outer cylinder rotate clockwise with an angular velocity $\Omega_O$, whereas the inner cylinder is clamped. Assuming an infinite length in the out-of-plane direction, we employ a two-dimensional model where velocities in the out-of-plane direction vanish. Each fluid particle is then characterized by an in-plane velocity vector and a pressure which depend only on its radial position. We assume the material to have either Newtonian or Oldroyd-B constitutive behavior. For these two material models, a stability analysis of the continuous Taylor-Couette model predicts instabilities in some ranges of problem parameters, see \cite{BAI2015} and references therein. Here we choose the parameters such that stability is preserved.

\begin{table}[tb]
\centering
\begin{tabular}{l l l} 
 \toprule
 parameter & notation & value\\
 \hline
 inner radius & $R_I\approx$ & $100$mm\\
 outer radius & $R_O\approx$ & $200$mm\\
 angular velocity & $\Omega_O$ & $7.5$s$^{-1}$\\
 \bottomrule
\end{tabular}
\caption{Parameters for the Taylor-Couette setup.}
\label{tab:TCparameters}
\end{table}

We choose a characteristic parent domain $\tilde{\Omega}=[0,10]$ and a parametric domain $\hat{\Omega}^n=\hat{\Omega}=[0,10]\times[0,2]$. The specific nature of Taylor-Couette flow requires the application of periodic boundary conditions to the parent domain. These act in characteristic direction where the modeling details are provided in \ref{periodicBCs}. At the initial time step, the boundary control points are distributed point symmetrically between inner radius $R^{CP}_I=100\mathrm{mm}$ and outer radius $R^{CP}_O=200\mathrm{mm}$. Different refinements are studied.

As level function we employ
\begin{equation}
\mathcal{L}(\bm{x}) = -10\cdot \dfrac{\varphi}{2\pi} = -\dfrac{10}{2\pi}\arctan{(y/x)},\ \ \ \ \ \ \ \ \nabla_{\bm{x}}\mathcal{L}(\bm{x})=\dfrac{10}{2\pi}
\begin{pmatrix}
\dfrac{y}{x^2+y^2}\\
-\dfrac{x}{x^2+y^2}
\end{pmatrix}
\end{equation}
This choice associates points along each radial line in the physical space to the same parametric level. A blending to ensure a rectangular parametric domain structure is not required due to the applied periodic boundary condition, hence $\gamma^n\left(\xi\right)=1$.

The imposition of boundary conditions is carried out first on all outer boundary control points of pressure and velocity, with index sets $\mathcal{A}_{outer}$ and $\mathcal{B}_{outer}$, respectively. Dirichlet boundary conditions are homogeneous for pressure, i.e. $q^n_a=0\mathrm{Pa} \ \forall\ a\in \mathcal{A}_{outer}$, and inhomogeneous for velocity
\begin{equation}
\bm{d}^n_b = 
\begin{pmatrix}
d^n_{xb}\\
d^n_{yb}\\
\end{pmatrix}
=\Omega_O
\begin{pmatrix}
c^n_{yb}\\
-c^n_{xb}\\
\end{pmatrix} \ \ \ \ \ \ \ \ \forall\ b\in \mathcal{B}_{outer}
\end{equation}
Inner boundary velocity control points with index set $\mathcal{B}_{inner}$ are subjected only to homogeneous Dirichlet boundary conditions on velocity, $\bm{d}^n_b=\bm{0}\mathrm{mm/s} \ \forall\ b\in \mathcal{B}_{inner}$.

The accuracy is quantified with the logarithm of the $L^2$ norm of the relative error in velocity and pressure with respect to the analytical reference solution. Computation of the reference results requires evaluation of the current radii $R_I$ and $R_O$. We evaluate these quantities at each time step from the current mesh boundary, as the radius of the outer boundary increases slightly during the simulation due to the incremental time stepping procedure. This effect is not due to FLIGA but to time integration and hence we compensate for it in our post-processing by recomputing the reference solution.

\subsection{Taylor-Couette flow: Newtonian fluid}

For this constitutive model, we adopt a solvent viscosity $\eta_s=50\mathrm{Pa\cdot s}$. \cite{BAI2015} provides the following analytical solution for the horizontal velocity profile
\begin{equation}
v_{Newt,x}(r,\varphi) = \sin(\varphi)\left(\alpha r+\dfrac{\beta}{r}\right)
\end{equation}
in which
\begin{equation}
\alpha = \dfrac{\Omega_O R_O^2 - \Omega_I R_I^2}{{R_O^2}-{R_I^2}}
\end{equation}
\begin{equation}
\beta = \dfrac{\left(\Omega_I - \Omega_O\right)R_I^2 R_O^2}{{R_O^2}-{R_I^2}}
\label{eq:beta}
\end{equation}
In our non-inertial Newtonian case, we do not expect pressure variations to occur and thus obtain for boundary condition $p_{Newt}(R_O)=0\mathrm{Pa}$
\begin{equation}
p_{Newt}(r) = 0\mathrm{Pa}
\end{equation}
We adopt $\Delta t=5.0\mathrm{e-5}\mathrm{s}$, however the time step does not have a significant influence on the results for the present setup.

As follows, we illustrate the numerical results obtained using FLIGA as well as classical IGA (Section \ref{sec:IGA}) for reference.
We first evaluate the $L^2$ norm of the relative error as a function of the deformation. We investigate linear, quadratic and cubic polynomial B-Spline orders, i.e. $r=1,2,3$, where the initial mesh consists of $18\times 6$ pressure and $36\times 12$ velocity elements. We initialize both IGA and FLIGA mesh with Gauss integration using $(r+1)\times (r+1)$ points.

\begin{figure}[h!]
\centering
\includegraphics[width=0.9\textwidth]{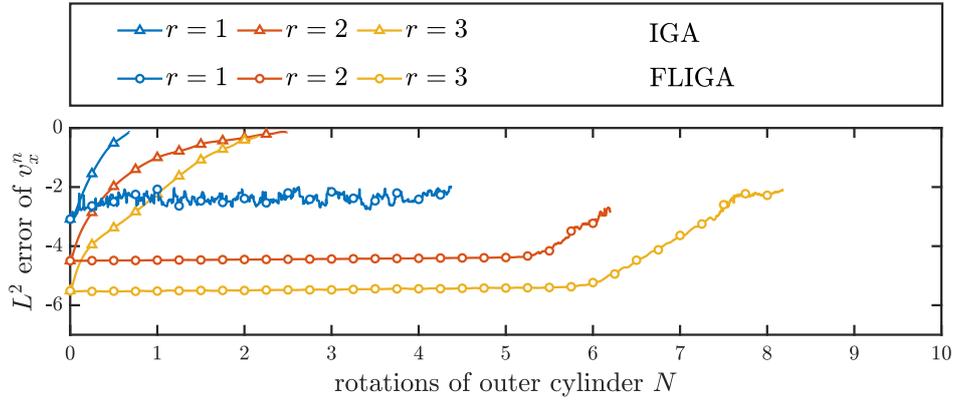}
\caption{Error over deformation in the Newtonian Taylor-Couette benchmark for polynomial degrees $r=1,2,3$.}
\label{fig:TCresultsNewtonian1}
\end{figure}

The results are reported in \autoref{fig:TCresultsNewtonian1}. It can be seen that the accuracy of IGA deteriorates at relatively small deformations, which is expected due to distortion of the classical mesh. Instead, with FLIGA the error stays nearly constant throughout the floating of the mesh for a remarkably large range of deformations. As the polynomial order $r$ increases, the accuracy of the results improves, however the computational cost also increases. The lack of smoothness is probably the reason for the lower accuracy obtained with linear basis functions ($r=1$) in combination with material point integration (note that quadrature points have to pass discontinuities in FLIGA).

\begin{figure}[h!]
\centering
\includegraphics[width=0.55\textwidth]{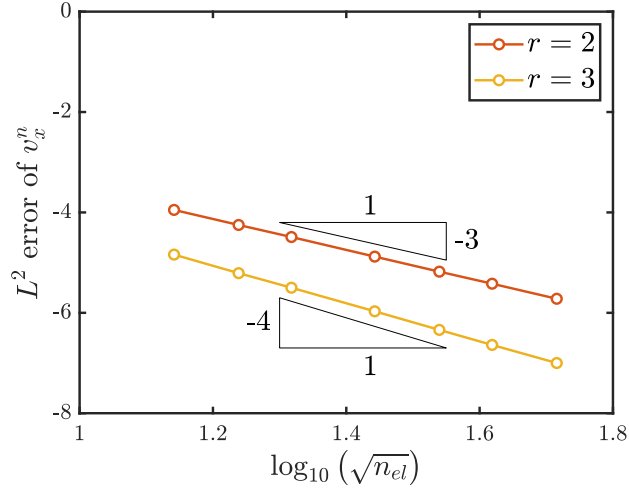}
\caption{Error plateau over spatial refinement in the Newtonian Taylor-Couette benchmark.}
\label{fig:TCresultsNewtonian2}
\end{figure}

The formation of an error plateau allows for an easy evaluation of convergence properties under mesh refinement. \autoref{fig:TCresultsNewtonian2} illustrates the  $L^2$ norm of the relative velocity error plateau for polynomial orders $r=2,3$  over spatial refinement. These results agree well with what is expected from classical IGA, as the floating mesh initially conforms with such a discretization, the convergence rate being close to $r+1$.

\FloatBarrier
\subsection{Taylor-Couette flow: Oldroyd-B fluid}

For the viscoelastic Taylor-Couette problem with the Oldroyd-B constitutive model, we choose a polymeric and a solvent viscosity of $\eta_p=150\mathrm{Pa\cdot s}$ and $\eta_s=50\mathrm{Pa\cdot s}$, respectively. The initial condition for the polymeric stress reads $\bm{\pi}^0=\bm{0}\mathrm{Pa}$ and the relaxation time is chosen as $\lambda=0.1\mathrm{s}$. Hence, we obtain a Weissenberg number of
\begin{equation}
Wi=\lambda\cdot \dot{\gamma}=\lambda\cdot \dfrac{\Omega_O R_O}{R_O-R_I}\approx 1.5
\end{equation}
characterizing the ratio of elastic to viscous forces in our benchmark flow. It is well known that most standard numerical schemes are only applicable to viscoelastic simulations up to $Wi\approx 1$ \citep{ZHANG2021}. Here, we maximize the elastic component by entering this regime, but do not investigate a further increase which is generally critical in the absence of stabilization.

As before, we compare FLIGA and IGA results with the analytical reference solution using the $L^2$ norm of the relative error. Note that in the viscoelastic case we have a transient stress response, which starts from the initial polymeric stress condition and approaches a steady state. We investigate only the steady state and use the reference solution in \citep{BAI2015}. The analytical steady state velocity field equals that of the Newtonian case, $v_{OldB,x}(r,\phi) = v_{Newt,x}(r,\phi)$. However, for the pressure boundary condition $p_{OldB}(R_O)=0\mathrm{Pa}$, we now expect a different pressure profile at steady state
\begin{equation}
p_{OldB}(r) =  2\beta^2 \lambda\pi \cdot \left(\frac{1}{r^4}-\frac{1}{R_O^4}\right)
\label{eq:pressuresolution}
\end{equation}
with $\beta$ given by \eqref{eq:beta}.

Once again we evaluate the $L^2$ norm of the relative error as a function of the deformation, where we now decrease the time increment to $\Delta t=1.0e-5\mathrm{s}$ for $r=1,2$, and $\Delta t=5.0e-6\mathrm{s}$ for $r=3$ due to the transient character of the problem. We adopt again $18\times 6$ pressure and $36\times 12$ velocity elements for the initial mesh and initialize material point integration on this mesh by distributing $(r+1)\times (r+1)$ Gauss points per element.

\begin{figure}[h!]
\centering
\includegraphics[width=0.9\textwidth]{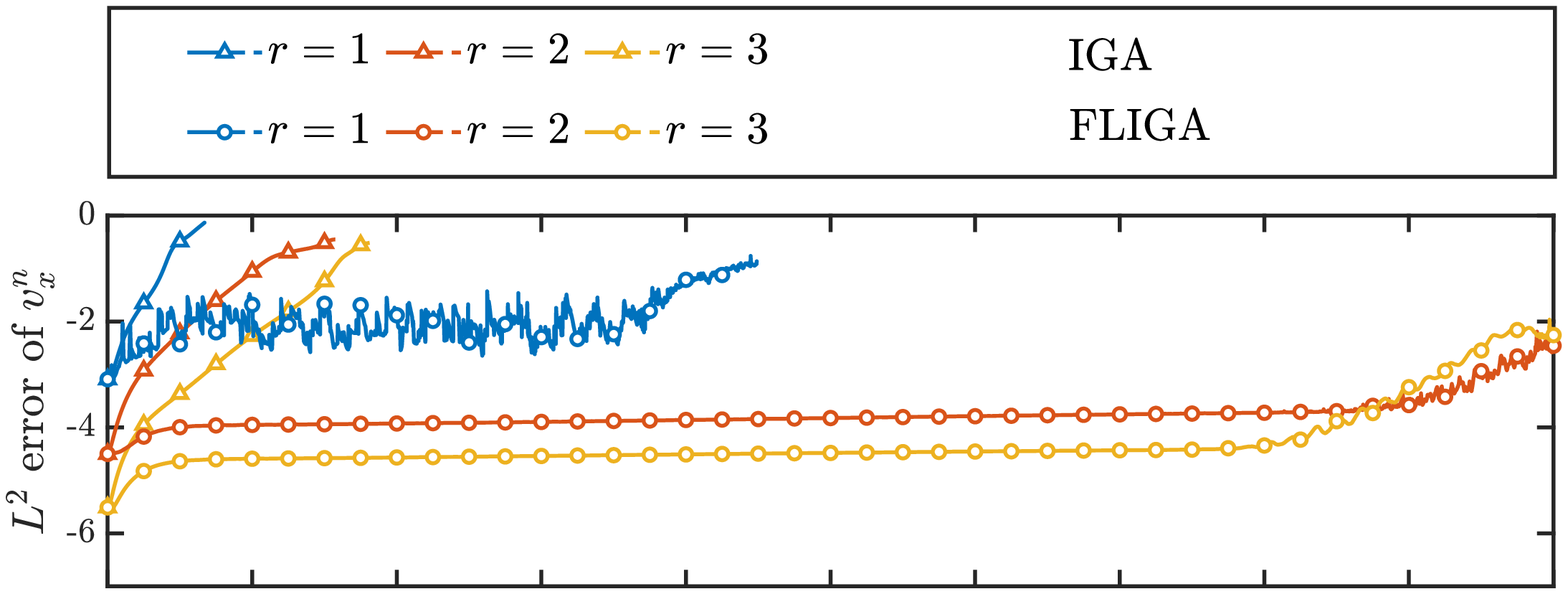}
\includegraphics[width=0.9\textwidth]{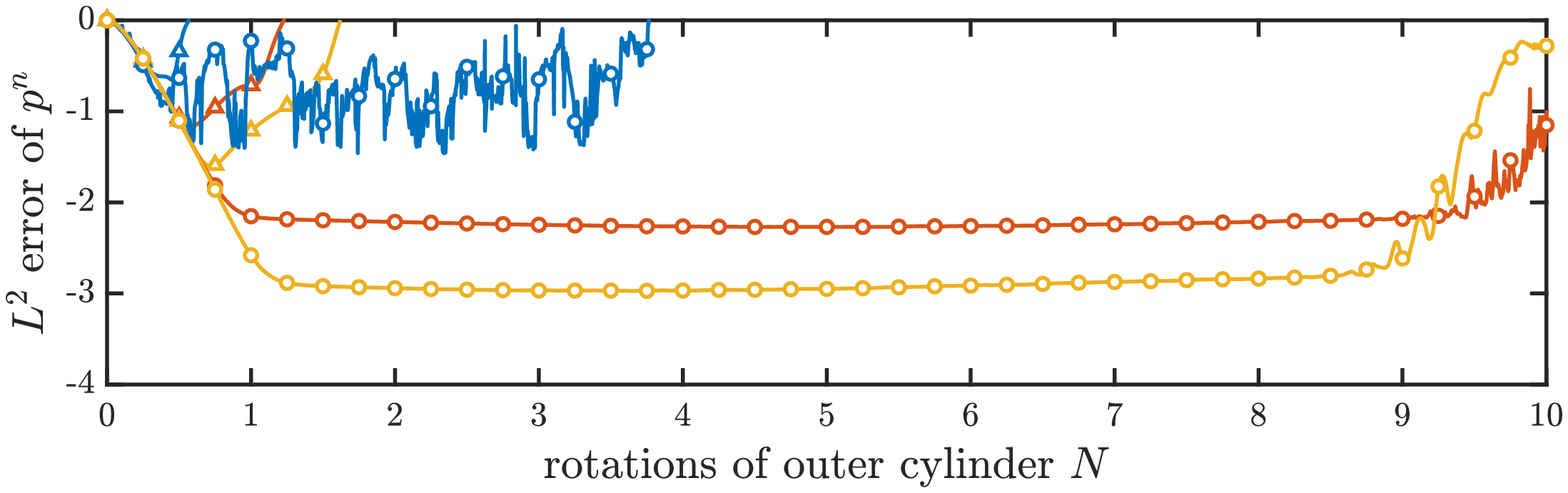}
\caption{Error over deformation in the viscoelastic Taylor-Couette benchmark for polynomial degrees $r=1,2,3$.}
\label{fig:TCresultsOldroydB1}
\end{figure}

\autoref{fig:TCresultsOldroydB1} shows at the top the error versus deformation for the horizontal velocity component. The error norm starts at a low level which corresponds to the Newtonian result since viscoelastic behavior is not yet active. As elastic contributions come into play, the accuracy slightly decreases and the error reaches a plateau at steady state. The effect of the polynomial degree on the accuracy agrees well with the previous observations in the Newtonian case. Surprisingly, for $r=2,3$ the extent of stable deformation is even significantly larger than in the Newtonian case.
At the bottom, \autoref{fig:TCresultsOldroydB1} displays the error versus deformation for the pressure field. As the analytical pressure given by \eqref{eq:pressuresolution} is valid only at steady state, the error is large in the initial state where, correctly, the numerical pressure is zero. Afterwards, again a plateau in the error is reached at steady state. Similar observations as for the velocity error hold in this case. 

\begin{figure}[h!]
\centering
\includegraphics[width=0.7\textwidth]{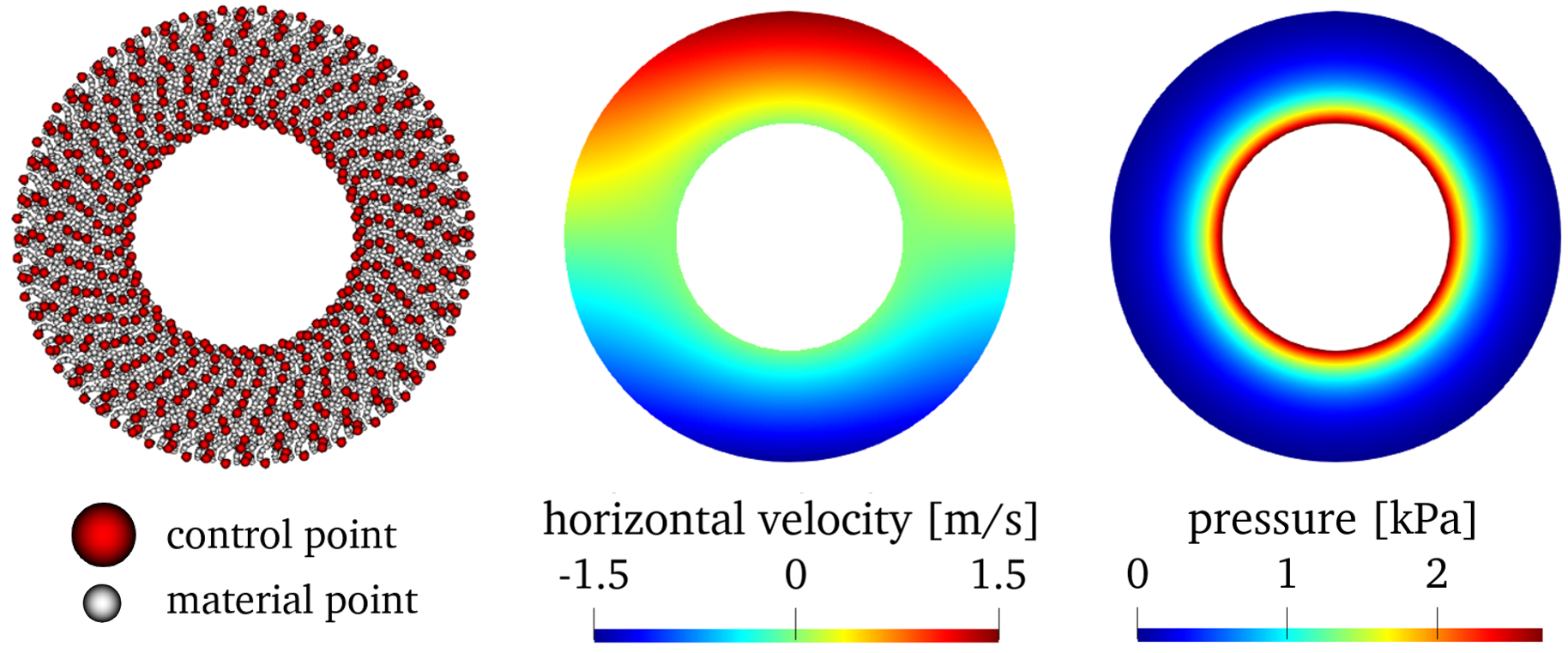}
\caption{Approximate horizontal velocity and pressure fields after $N=2$ rotations of the outer cylinder.}
\label{fig:TCresultsOldroydB2}
\end{figure}

The approximate velocity and pressure fields for $r=3$ after $N=2$ rotations of the outer cylinder are shown in \autoref{fig:TCresultsOldroydB2}. Both fields are free from oscillations, which is an indication of the absence of locking effects.

\begin{figure}[h!]
\centering
\includegraphics[width=0.55\textwidth]{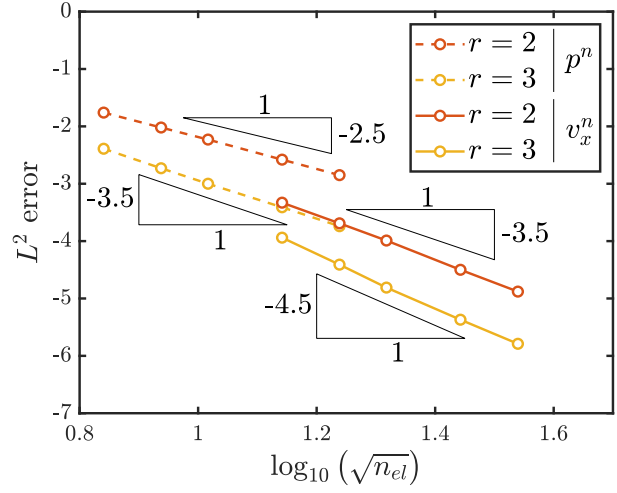}
\caption{Error plateau for the steady state of pressure and velocity over spatial refinement in the Oldroyd-B Taylor-Couette benchmark.}
\label{fig:TCresultsOldroydB3}
\end{figure}

We conduct another mesh refinement study for $r=2,3$, see \autoref{fig:TCresultsOldroydB3}. The parameters are the same as before, but the time step is adjusted for fine floating meshes to guarantee that the error is controlled by the spatial discretization. We obtain for both velocity and pressure a convergence rate that even exceeds the expected order of $r+1$ and $r$, respectively.

\FloatBarrier

\begin{figure}[h]
\centering
\includegraphics[trim={0cm 0cm 0cm 0cm},clip,width=0.475\textwidth]{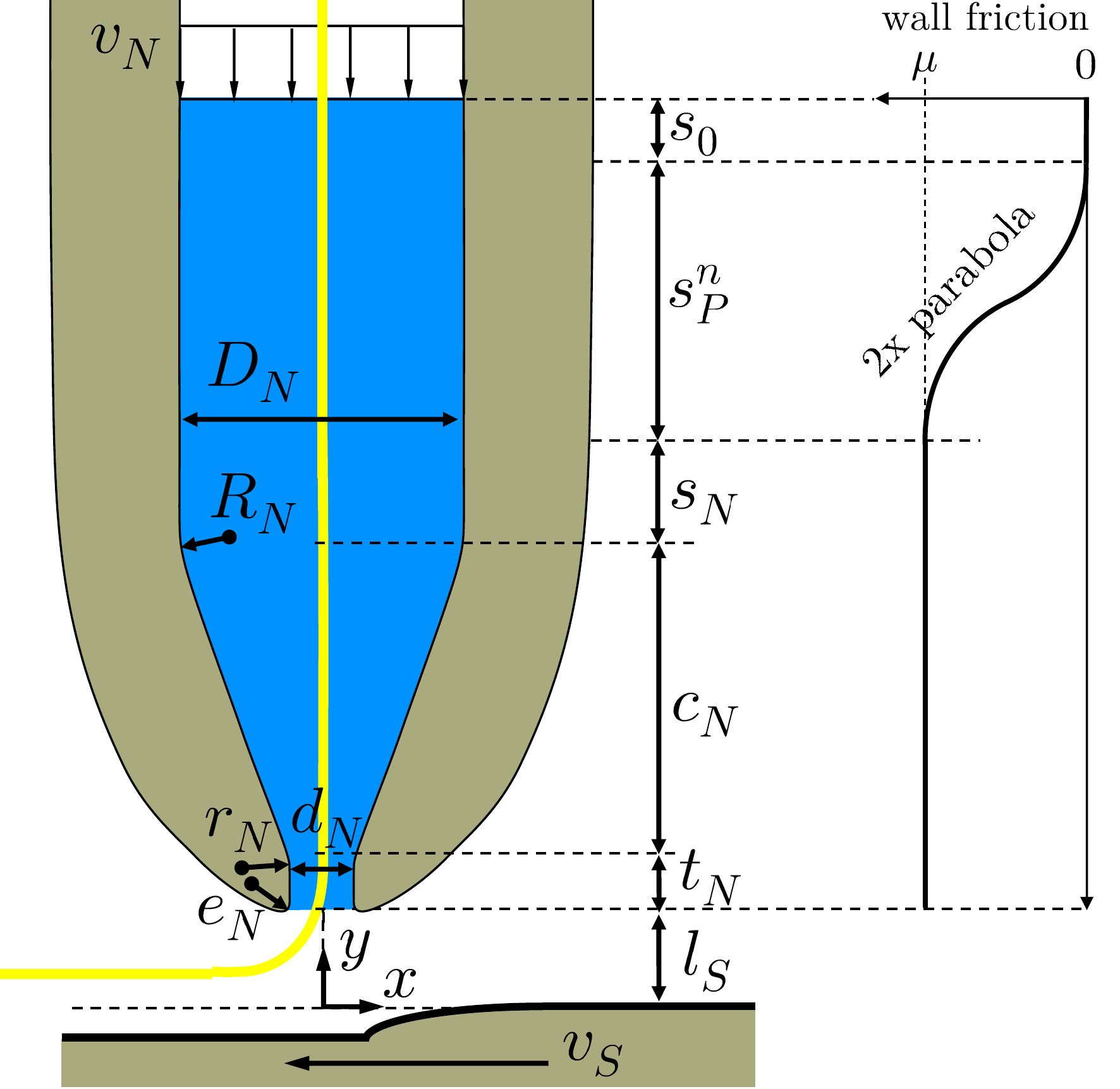}
\caption{Schematic extrusion setup.}
\label{fig:AMsetup}
\end{figure}

\subsection{Extrusion-based AM}

In this third numerical example, we demonstrate the suitability of FLIGA for the simulation of polymer extrusion processes in AM, which was our original motivation for FLIGA.
The problem setup consists of an extrusion nozzle, that is initially filled with material, see \autoref{fig:AMsetup} for details. The contact between material and nozzle walls is described by a node-to-surface frictional contact formulation (Coulomb friction), where the contact constraints are enforced with the penalty method. To facilitate the employment of a plug flow boundary condition driving the extrusion, the friction factor fades in from no wall friction at the upper domain area ($s_0$) to a plateau. This is modeled by two equal-sized sections of parabolic increase, see again the figure for more details. Note that we keep all measures describing the nozzle constant throughout the simulation except the transition area for wall friction, $s_P^n$, which is determined by the current coordinate level of the top boundary. Values for wall penalty factor, plateau friction coefficient as well as the other parameters are given in \autoref{tab:AMparameters}.

Additionally, we impose on the two top control point rows ($i=1,2$) the condition
\begin{equation}
d^n_{x(i)(j)}=d^{n-1}_{x(\alpha)(j)}
\end{equation}
with $\alpha=11$, which has a stabilizing effect on the horizontal deformation in the top region and only negligibly affects the overall simulation result.

We employ the Oldroyd-B viscoelastic model with initial condition for the polymeric stress $\bm{\pi}^0=\bm{0}\mathrm{Pa}$. Once the extrudate reaches below a threshold vertical coordinate, it attaches irreversibly to a substrate modeled by Dirichlet boundary conditions. The substrate level is adapted as to avoid a singularity in the physical mapping at points associated to the corners of the parametric domain (see \autoref{fig:AMsetup}). As inertia forces are absent from the model, instead of moving the nozzle, we can impose the printing movement on the substrate \citep{COMMINAL2018}.

\begin{table}[tb]
\centering
\begin{tabular}{l l l} 
 \toprule
 parameter & notation & value\\
 \hline
 inlet area & $s_0$ & $0.25$mm\\
 parabolic friction area at $t=0$ & $s_P^0$ & $5.15$mm\\
 straight section & $s_N$ & $1.1$mm\\
 contraction area & $c_N$ & $2.0$mm\\
 nozzle tip & $t_N$ & $0.8$mm\\
 substrate distance & $l_S$ & $0.525$mm\\
 inlet diamater & $D_N$ & $2.0$mm\\
 outlet diameter & $d_N$ & $0.5$mm\\
 convergent radius & $R_N$ & $2.5$mm\\
 divergent radius & $r_N$ & $1.2$mm\\
 nozzle exit radius & $e_N$ & $0.3$mm\\
 inlet velocity & $v_N$ & $10.0$mm/s\\
 substrate speed & $v_S$ & $38.0$mm/s\\
 wall penalty factor& $k$ & $1.0$e$8$\\
 plateau friction coefficient & $\mu$ & $0.075$\\
 relaxation time & $\lambda$ & $0.04$s\\
 polymeric viscosity & $\eta_p$ & $1.5$kPa$\cdot$ s\\
 solvent viscosity & $\eta_s$ & $0.05$kPa$\cdot$ s\\
 \bottomrule
\end{tabular}
\caption{Parameters for the extrusion setup.}
\label{tab:AMparameters}
\end{table}

A time step of $\Delta t=2.0\mathrm{e}-5\mathrm{s}$ is selected. Spatial discretization starts with the definition of the polynomial order $r=2$ for both velocity and pressure. In the initial configuration of a classical mesh, we employ $4$ pressure and $8$ velocity elements per cross-section and use a large number of elements in extrusion direction ($480$ for pressure and $960$ for velocity). The parent and initial parametric domain read $\tilde{\Omega}=\left[0,9\right]$ and $\hat{\Omega}^0=\left[0,9\right]\times\left[-1,1\right]$, respectively. We choose material point integration with $8\times 3$ Gauss points per parametric element in the initial configuration.

We employ the following blending function
\begin{equation}
\gamma(\xi) =
\begin{cases}
0 & \xi = a^n \\
\text{lin. ramp} & a^n < \xi \leq a^n + 2d \\
1 & a^n + 2d < \xi \leq b^n - 2d \\
\text{lin. ramp} & b^n - 2d < \xi \leq b^n - \phantom{2}d \\
0 & b^n - \phantom{2}d < \xi \\
\end{cases}
\end{equation}
with $d=1.0\mathrm{mm}$. The level function is chosen such that
\begin{equation}
\begin{split}
\mathrm{before\ attachment:}\ \ \ &\nabla_{\bm{x}}\mathcal{L}(\bm{x})=
\begin{cases}
\begin{pmatrix}
0\\
-1
\end{pmatrix}
\end{cases}\\
\mathrm{once\ attached:}\ \ \ &\nabla_{\bm{x}}\mathcal{L}(\bm{x})=
\begin{cases}
\begin{pmatrix}
0\\
-1
\end{pmatrix} &P_{Ry}\leq y \\
\begin{pmatrix}
\phantom{-}r_y/\sqrt{\bm{r}\cdot\bm{r}}\\
-r_x/\sqrt{\bm{r}\cdot\bm{r}}
\end{pmatrix} &y<P_{Ry}, \ P_{Rx}\leq x \\
\begin{pmatrix}
-1\\
0
\end{pmatrix} &y<P_{Ry}, \ \ x<P_{Rx} \\
\end{cases}
\end{split}
\end{equation}
where $\bm{r}=\bm{x}-\bm{P}_R$, with $\bm{P}_R=\left(-\dfrac{d_N}{2}-e_N-0.1\ ,\ l_s+e_N+0.1\right)^T$ for the coordinate system defined in \autoref{fig:AMsetup}. This choice defines a level function as qualitatively illustrated in \autoref{fig:AM_Levelcurves}.

\begin{figure}[b!]
\centering
\includegraphics[width=\textwidth]{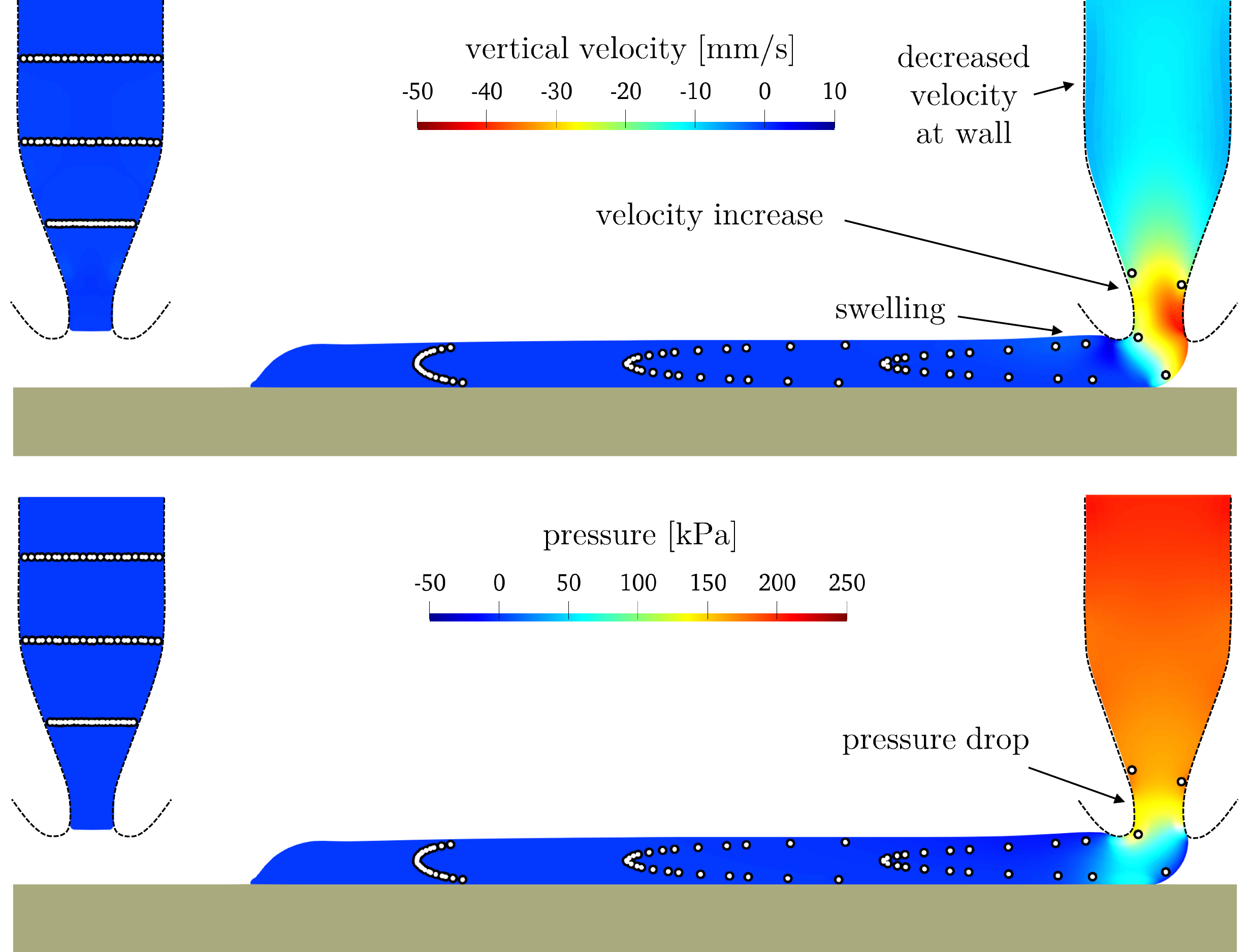}
\caption{Extrusion simulation with FLIGA. The white dots highlight material points which were horizontally aligned at the initial step.}
\label{fig:AM_Simulation1}
\end{figure}

\autoref{fig:AM_Simulation1} shows the obtained solution fields prior to deformation and during extrusion at time $T\approx 0.4s$. The upper part of the figure shows the vertical velocity field. At a given horizontal level in the deformed state, the absolute velocity at the nozzle walls is always observed to be lower than at the center of the nozzle, which is due to wall friction. The significant increase of vertical velocity along the nozzle contraction is due to the incompressibility constraint. Also, we observe viscoelastic swelling after the polymer leaves the confinement of the nozzle. In the bottom part of the figure, the pressure field is shown. This is free from oscillations, indicating the absence of locking. Further, the pressure in the deformed state decreases along the contraction zone and finally becomes close to zero at the nozzle exit. This behavior is consistent with observations in the AM literature, see e.g. \cite{TURNER2014}. 

The position of a selection of white dots aligned horizontally at the initial time step is tracked graphically to qualitatively illustrate the extreme deformations occurring during extrusion, which are successfully handled by FLIGA.

\FloatBarrier

\section{Conclusions}
\label{sec:conclusions}

We propose Floating Isogeometric Analysis (FLIGA) as a novel extension of classical IGA. FLIGA enables the solution of problems featuring extreme deformations which occur predominantly along one characteristic axis. The basic idea is a modification of the classical tensor-product structure of B-Spline basis functions, such that multiple B-Spline chains in the characteristic direction are able to independently float against each other in the parametric domain. Following this strategy, distortion introduced by the physical mapping is overcome without the need for classical remeshing. FLIGA enjoys several benefits of isogeometric and meshless modeling: (i) the initial geometry representation is a classical IGA B-Spline mesh preserving the isogeometric principle, (ii) several properties of B-Splines favorable for analysis are preserved e.g., partition of unity, weak Kronecker-delta property and first-order consistency, (iii) moving domain boundaries are naturally dealt with, (iv) mesh distortion is overcome, (v) we can easily construct mixed approximation spaces to avoid locking due to e.g. incompressibility, (vi) there is no need to stabilize advective terms due to the Lagrangian quadrature point character. The effectiveness and accuracy of FLIGA were demonstrated and quantified with the benchmark problem of Taylor-Couette flow with Newtonian and viscoelastic Oldroyd-B material models. Accurate results were obtained up to extreme deformation levels while keeping the spatial convergence rate expected for IGA. Finally, FLIGA was applied to the viscoelastic simulation of extrusion-based additive manufacturing, showing stable and qualitatively accurate results. 

\appendix

\section{Newton-Raphson scheme for solving the governing system of equations}
\label{globalNewtonRaphson}

To compactly formulate the global Newton-Raphson scheme, we summarize at the current time step all unknowns in a single vector $\bm{u}$ and all residual components in a single vector $\bm{R}$. We introduce $\bm{K}$ as tangent stiffness matrix, i.e. the linearization of the residual vector w.r.t. to the unknowns. For brevity, we omit here the time step index $n$. With

\begin{equation}
\bm{u}=
\begin{bmatrix}
q_a\\
d_{bi}\\
\end{bmatrix}\ \ \ \ \ \ \ \ \ \ \ \ \ \ 
\bm{R}=
\begin{bmatrix}
Q_a\\
S_{bi}\\
\end{bmatrix}\ \ \ \ \ \ \ \ \ \ \ \ \ \ 
\bm{K}= \dfrac{\partial \bm{R}}{\partial \bm{u}} =
\begin{bmatrix}
\dfrac{\partial Q_a}{\partial q_{\alpha}} & \dfrac{\partial Q_a}{\partial d_{\beta k}}\\
\dfrac{\partial S_{bi}}{\partial q_{\alpha}} & \dfrac{\partial S_{bi}}{\partial d_{\beta k}}\\
\end{bmatrix}
\label{eq:stiffnesstangent}
\end{equation}
the contributions to the tangent stiffness matrix read as follows
\begin{equation}
\dfrac{\partial Q_a}{\partial q_{\alpha}} = 0\ \ \ \ \ \ \ \ \ \ \ \ \ \ 
\dfrac{\partial Q_a}{\partial d_{\beta k}} = \int_{\Omega} \dfrac{\partial B_{\beta}}{\partial x_k} A_a\ \mathrm{d}\Omega
\end{equation}

\begin{equation}
\dfrac{\partial S_{bi}}{\partial q_{\alpha}} = \int_{\Omega} \left( -A_{\alpha} \delta_{ij} + \dfrac{\partial\tau_{ij}}{\partial q_{\alpha}}\right) \dfrac{\partial B_b}{\partial x_j} \ \mathrm{d} \Omega\ \ \ \ \ \ \ \ \ \ \ \ \ \ 
\dfrac{\partial S_{bi}}{\partial d_{\beta k}} = \int_{\Omega} \dfrac{\partial\tau_{ij}}{\partial d_{\beta k}} \dfrac{\partial B_b}{\partial x_j} \ \mathrm{d} \Omega
\end{equation}
The linearization of the deviatoric stress is given in \ref{deviatoriccauchylinearizations}.

The Newton-Raphson scheme solves the nonlinear vector equation
\begin{equation}
\bm{R}\left(\bm{u}\right)=\bm{0}
\end{equation}
by the iterative formula
\begin{equation}
\bm{u}^{(r+1)}=\bm{u}^{(r)}-\bm{K}^{-1}\bm{R}\left(\bm{u}^{(r)}\right)
\end{equation}
Starting from the initial guess $\bm{u}^{(0)}$ (that can be obtained from the previous time step) we repeat application of this formula until $\Vert \bm{R}\left(\bm{u}^{(r^*)}\right)\Vert<\mathrm{tol}$ and thus identify $\bm{u}\leftarrow \bm{u}^{(r^*)}$. We choose $\mathrm{tol}=1.0e-9$.

\section{Newton-Raphson scheme for the determination of the parent quadrature point position}
\label{parentnewton}
In order to determine $\tilde{\xi}_{qj}$ (time step index $n$ omitted) as defined by \autoref{eq:inverseG}, we introduce the nonlinear scalar equation
\begin{equation}
R_j\left(\tilde{\xi}\right)=\mathcal{G}_j\left(\tilde{\xi}\right)-\xi_q
\label{eq:parentnewton}
\end{equation}
based on \autoref{eq:Gmapping} with the unknown parent coordinate being the root $\tilde{\xi}_{qj}: \ R_j\left(\tilde{\xi}_{qj}\right)=0$.
Given the parametric quadrature point coordinate $\xi_q$ and $J_j$ as the Jacobian of $\mathcal{G}_j$ (\autoref{eq:Gjacobian}), we now apply the Newton-Raphson scheme
\begin{equation}
\tilde{\xi}^{(r+1)}_{qj}=\tilde{\xi}^{(r)}_{qj}-J_j\left(\tilde{\xi}^{(r)}_{qj}\right)^{-1}R_j\left(\tilde{\xi}^{(r)}_{qj}\right)
\end{equation}
by starting from the initial guess $\tilde{\xi}^{(0)}_{qj}$ (obtained from the previous time step) until at iteration $r^*: \Vert R_j\left(\tilde{\xi}^{(r^*)}_{qj}\right)\Vert<\mathrm{tol}\approx 1\mathrm{e}-10$. Finally, we obtain $\tilde{\xi}_{qj}\leftarrow \tilde{\xi}^{(r^*)}_{qj}$.

Note that an initial guess might not be available when a material point has just crossed an element boundary in normal direction. In this case, we can solve \autoref{eq:parentnewton} once by the bisection method and then apply the proposed Newton scheme again for the subsequent time steps.

\section{Continuity}
\label{continuityandsupport}

\begin{figure}[t]
\centering
\includegraphics[trim={5.25cm 2cm 4.25cm 1.5cm}, clip,width=0.55\textwidth]{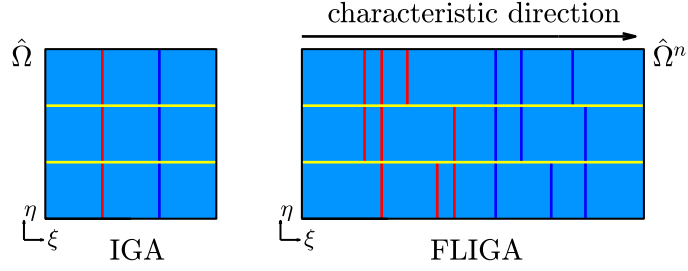}
\caption{Schematic visualization of parametric lines with limited continuity of B-Spline approximations by IGA and FLIGA.}
\label{fig:continuity}
\end{figure}

Basis function continuity in FLIGA is exemplarily illustrated on the parametric domain for the arbitrary floating of an IGA mesh with $r=2$ and $3\times 3$ elements in \autoref{fig:continuity}. For both IGA and FLIGA, the continuity of B-Spline approximations along the depicted colored lines is $C^{r-1}$, assuming equal polynomial order of parent and normal bases and no repetition of inner knots in the respective knot vectors. In the interior blue areas continuity is $C^{\infty}$. It is observed that less discontinuity lines of the single B-Splines are aligned in the case of FLIGA.

\section{Partition of unity}
\label{partitionofunity}

We show as follows that FLIGA B-Splines satisfy partition of unity at any point $\bm{x}_q \in \Omega$, i.e.
\begin{equation}
\sum_i \sum_j B_{ij}\left(\bm{x}_q\right)=1
\label{eq:PU1}
\end{equation}
For brevity, we omit the time step index $n$.

\noindent \textit{Proof:}

\noindent Let us reformulate the condition employing \eqref{eq:physicalbsplinesevaluated} and \eqref{eq:parametricbsplinesevaluated}
\begin{equation}
1=\sum_i \sum_j \tilde{N}_{i}\left(\tilde{\xi}_{qj}\right)\hat{M}_{j}\left(\eta_q\right)
\label{eq:PU1}
\end{equation}
The latter requirement can also be expressed as
\begin{equation}
1=\sum_j \left( \hat{M}_{j}\left(\eta_q\right) \sum_i \tilde{N}_{i}\left(\tilde{\xi}_{qj}\right) \right)
\end{equation}
Note that, since $\tilde{\xi}_{qj}$ is invariant w.r.t. $i$ and since the parent and normal bases fulfil partition of unity (\cite{PiegTill96}), it holds
\begin{equation}
\begin{split}
&\sum_i \tilde{N}_{i}\left(\tilde{\xi}_{qj}\right)=1\ ,\ \ \ \ \ \ \ j=\mathrm{const}\\
&\sum_j\hat{M}_{j}\left(\eta_q\right)=1
\end{split}
\end{equation}
This finally leads to
\begin{equation}
1=\sum_j \left( \hat{M}_{j}\left(\eta_q\right) \sum_i \tilde{N}_{i}\left(\tilde{\xi}_{qj}\right) \right) = \sum_j \left( \hat{M}_{j}\left(\eta_q\right) \cdot 1\right)=1
\end{equation}
proving partition of unity for FLIGA B-Spline bases.

\section{Weak Kronecker-delta property}
\label{weakkroneckerdeltaproperty}

FLIGA B-Spline bases constructed on the analysis domain $\Omega$ possess another important feature, namely weak Kronecker-delta property
\begin{equation}
B_{ij}\left(\bm{x}_q\right) = 0 \ \ \ \ \ \ \ \ \forall \ \bm{x}_q \in \partial\Omega,\ (i,j) \notin \mathcal{B}_{bound}
\label{eq:weakkronecker1}
\end{equation}
where $\mathcal{B}_{bound}$ denotes the set of basis function index tuples associated to boundary control points and where again we drop index $n$.

\noindent \textit{Proof:}

\noindent Let us apply \eqref{eq:physicalbsplinesevaluated} and \eqref{eq:parametricbsplinesevaluated} such that proving
\begin{equation}
\tilde{N}_{i}\left(\tilde{\xi}_{qj}\right)\hat{M}_{j}\left(\eta_q\right) = 0 \ \ \ \ \ \ \ \ \forall \ \bm{\xi}_q \in \partial\Omega,\ (i,j) \notin \mathcal{B}_{bound}
\label{eq:weakkronecker2}
\end{equation}
is equivalent to prove \eqref{eq:weakkronecker1}. The (floating) tensor-product structure allows index-based determination of interior control points by
\begin{equation}
(i,j) \notin \mathcal{B}_{bound} \Leftrightarrow (i\neq1) \wedge (i\neq I) \wedge (j\neq 1) \wedge (j\neq J)
\label{eq:weakkroneckerconstraint}
\end{equation}
which is similar to classical IGA.
Further, due to the rectangular structure of the parametric domain, we can guarantee
\begin{equation}
\begin{split}
\bm{\xi}_q \in \partial\Omega \Leftrightarrow\ &(\xi_q = a) \vee (\xi_q = b) \vee (\eta_q = \eta_1) \vee (\eta_q = \eta_{J+r+1})\ , \ \mathrm{while}\\
&(\xi_q = a) \Leftrightarrow (\tilde{\xi}_{qj} = \tilde{\xi}_1\ \forall j=1,...,J)\\
&(\xi_q = b) \Leftrightarrow (\tilde{\xi}_{qj} = \tilde{\xi}_{I+r+1}\ \forall j=1,...,J)
\end{split}
\label{eq:fourcases}
\end{equation}
Evaluation of \eqref{eq:weakkronecker2} can now be carried out separately for the four possible cases represented in \eqref{eq:fourcases}. Let us assume that we are in the first case, i.e. $\xi_q = a$ or $\tilde{\xi}_{qj} = \tilde{\xi}_1\ \forall j=1,...,J$ while due to \eqref{eq:weakkroneckerconstraint} $i\neq1$. This implies an assessment of weak Kronecker-delta property for the parent basis, which is a classical 1D B-Spline basis. Fulfillment was shown in \cite{PiegTill96}. Satisfying $i\neq 1$, we thus obtain $\tilde{N}_{i}\left(\tilde{\xi}_{qj}\right)=\tilde{N}_{i}\left(\tilde{\xi}_1\right)=0$ resulting in $\hat{B}_{ij}\left(\bm{\xi}_q\right) = 0$ for the case at hand, see \eqref{eq:weakkronecker2}. With this argumentation, we demonstrated weak Kronecker-delta property of FLIGA bases at the left parametric domain boundary $\xi=a$.

The other three cases of \eqref{eq:fourcases} correspond to the remaining three edges and can be handled similarly. As each case satisfies \eqref{eq:weakkronecker2} (condition \eqref{eq:weakkroneckerconstraint} fulfilled) there is no option left for which the weak Kronecker-delta property could be violated. Hence, all four parametric edges exhibit the weak Kronecker-delta property. Indeed, this property can be recognized from \autoref{fig:fligaconcept} to follow from the weak Kronecker-delta property of the characteristic and normal bases.

\section{Boundary preservation}
\label{boundary preservation}

Another advantageous feature of FLIGA B-Spline bases is the independence of $\partial\Omega$ from the floating regulation points $h_{ij}$ (time step index $n$ skipped). That is, the physical boundary is not altered by the floating of characteristic basis functions, but only by the mechanical evolution of the problem.

\noindent \textit{Proof:}

\noindent To prove this, let us consider a point $\tilde{\xi}^*_{q1} \in \tilde{\Omega}$, where index $*$ indicates that this point is fixed in the parent domain. Note that this parent point is associated only to the floating map $j=1$. Let us now map this point to the parametric domain.
\begin{equation}
\xi_q = \mathcal{G}_1\left( \tilde{\xi}^*_{q1} \right) = \sum_{i} h_{i1}\tilde{N}_{i}\left(\tilde{\xi}^*_{q1}\right)
\label{eq:boundarypreservation1}
\end{equation}
Obviously, $\xi_q$ depends on the floating regulation points. However, let us further consider the physical mapping of the parametric point $\bm{\xi}_q=\left(\xi_q,\eta_1\right)^T$, where we associate our previously determined $\xi_q$ to the bottom boundary $\eta=\eta_1$. With partition of unity and weak Kronecker-delta property of the normal basis, as well as \eqref{eq:Fmapping},\eqref{eq:parametricbsplinesevaluated},\eqref{eq:CharacteristicBases} and \eqref{eq:boundarypreservation1}, we have all required information to compute the associated point $\bm{x}_q$ on the physical boundary.
\begin{equation}
\bm{x}_q = \sum_{i}\sum_{j} \bm{c}_{ij}\hat{B}_{ij}\left(\bm{\xi}_q\right) = \sum_{i}\sum_{j} \bm{c}_{ij}\hat{N}_{ij}(\xi_q)\hat{M}_j(\eta_1) = \sum_{i} \bm{c}_{i1}\hat{N}_{i1}(\xi_q) = \sum_{i} \bm{c}_{i1}\tilde{N}_{i}\left(\mathcal{G}^{-1}_1\left(\xi_q\right) \right) = \sum_{i} \bm{c}_{i1}\tilde{N}_{i}\left(\tilde{\xi}^*_{q1} \right)
\end{equation}
Unlike the characteristic $\xi_q$, the physical $\bm{x}_q$ proves not to be dependent on $h_{ij}$. Recall that $\tilde{\xi}^*_{q1}$ is prescribed constant and therefore $\tilde{N}_{i}\left(\tilde{\xi}^*_{q1} \right)$ remains constant as well. Following this demonstration for each $\tilde{\xi}^*_{q1} \in \tilde{\Omega}$ while $\eta=\eta_1$, the related $\bm{x}_q$ cover the entire physical boundary associated to the bottom edge of the parametric domain. Dependency is only w.r.t. $\bm{c}_{ij}$, hence, the physical shape is preserved under variation of $h_{ij}$. The same argumentation is valid for the upper edge of the parametric domain $\eta=\eta_{J+r+1}$.

Due to the weak Kronecker-delta property, the physical maps of the left and right edges of the parametric domain ($\xi=a$ and $\xi=b$) are fully determined by $\bm{c}_{ij}$ (with $i=1$ and $i=I$, respectively) and the normal basis $\left\lbrace\hat{M}_j\right\rbrace^{J}_{j=1}$. The physical shape of these edges is naturally preserved under floating, due to the normal basis remaining static under variation of $h_{ij}$.

\section{Newton-Raphson scheme for the determination of the parametric material point position}
\label{materialpointnewtonraphson}
Here, we focus on how to map back a physical point $\bm{x}_p$ to parametric space with geometry mapping $\mathcal{F}$ as given in \eqref{eq:Fmapping}. For simplicity we omit superscript $n$.
\begin{equation}
\bm{\xi}_q = \mathcal{F}^{-1}\left(\bm{x}_q\right)
\end{equation}
Analogous to the 1D case in \autoref{eq:parentnewton}, we apply a Newton-Raphson scheme on the nonlinear vector equation
\begin{equation}
\bm{R}\left(\bm{\xi}\right)=\mathcal{F}\left(\bm{\xi}\right)-\bm{x}_q
\end{equation}
for which we seek the root $\bm{\xi}_q$ such that $\bm{R}\left(\bm{\xi}_q\right)=\bm{0}$. Therefore, the linearization of $\mathcal{F}$ w.r.t. $\bm{\xi}$ is needed, which is known as Jacobian $\bm{J}$ by \eqref{eq:Fjacobian}.

Starting from an initial guess $\bm{\xi}_q^{(r=0)}$ (obtained from the previous time step), we apply
\begin{equation}
\bm{\xi}_q^{(r+1)}=\bm{\xi}_q^{(r)}-\bm{J}\left(\bm{\xi}_q^{(r)}\right)^{-1}\bm{R}\left(\bm{\xi}_q^{(r)}\right)
\end{equation}
until $\Vert\bm{R}\left(\bm{\xi}_q^{(r^*)}\right)\Vert<\mathrm{tol}$. We choose the same tolerance as for the determination of parent quadrature point positions $(\mathrm{tol}=1\mathrm{e}-10)$ and then set $\bm{\xi}_q\leftarrow \bm{\xi}^{(r^*)}_q$.

At each step $r$ of this scheme, we need to provide $\hat{B}_{ij}\left(\bm{\xi}_q^{(r)}\right)$ and its gradient $\nabla_{\bm{\xi}}\hat{B}_{ij}\left(\bm{\xi}_q^{(r)}\right)$ where $\bm{\xi}_q^{(r)}=(\xi_q^{(r)},\eta_q^{(r)})^T$ is given, therefore we follow the basis function evaluation strategy proposed in Section \ref{ssec:evaluation}.

\section{Deviatoric Cauchy stress linearizations} \label{deviatoriccauchylinearizations}
\begin{equation}
\dfrac{\partial\tau_{s,ij}}{\partial q_{\alpha}} = \dfrac{\partial\tau_{OldB,ij}}{\partial q_{\alpha}} = 0\ \ \ \ \ \ \ \ \ \ \ \ \ \ 
\dfrac{\partial\tau_{s,ij}}{\partial d_{\beta k}} = \dfrac{\partial\tau_{OldB,ij}}{\partial d_{\beta k}} = \eta_s \left(\dfrac{\partial B_{\beta}}{\partial x_j} \delta_{ik} + \dfrac{\partial B_{\beta}}{\partial x_i} \delta_{jk} \right)
\end{equation}

\section{Stiffness tangent matrix for implementation} \label{finalstiffnesstangent}

The final tangent stiffness matrix according to \eqref{eq:stiffnesstangent} can be implemented as:
\begin{equation}
\dfrac{\partial Q_a}{\partial q_{\alpha}} = 0\ \ \ \ \ \ \ \ \ \ \ \ \ \ 
\dfrac{\partial Q_a}{\partial d_{\beta k}} = \sum_q W_q\cdot \left.\left(\dfrac{\partial B_{\beta}}{\partial x_k}\cdot A_a\right)\right|_{\bm{x}=\bm{x}_q}
\end{equation}

\begin{equation}
\dfrac{\partial S_{bi}}{\partial q_{\alpha}} =  -\sum_q W_q\cdot \left.\left(\dfrac{\partial B_b}{\partial x_i}\cdot A_{\alpha} \right)\right|_{\bm{x}=\bm{x}_q}\ \ \ \ \ \ \ \ \ \ \ \ \ \ 
\dfrac{\partial S_{bi}}{\partial d_{\beta k}} = \eta_s \sum_q W_q\cdot \left.\left[ \left( \sum_j\dfrac{\partial B_{\beta}}{\partial x_j}\dfrac{\partial B_b}{\partial x_j}\right)\delta_{ik}+ \dfrac{\partial B_{\beta}}{\partial x_i}\dfrac{\partial B_b}{\partial x_k} \right]\right|_{\bm{x}=\bm{x}_q}
\end{equation}

\section{Periodic bases}
\label{periodicBCs}
\begin{figure}[h]
\centering
\includegraphics[trim={2cm 2.5cm 2cm 1.5cm},clip,width=12cm]{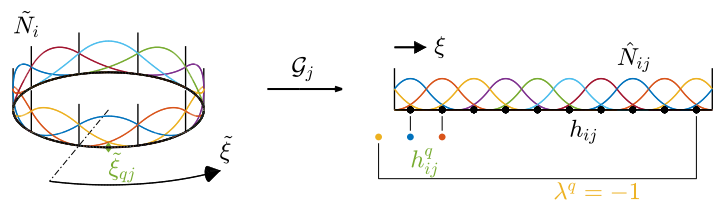}
\caption{Schematic overview of 1D periodic B-Spline basis construction.}
\label{fig:fligaconceptPBC}
\end{figure}

For certain problems, periodic boundary conditions in the characteristic direction are required. The parent B-Spline basis of order $r$ is therefore constructed on a closed 1D parent domain $\tilde{\Omega}$ and whenever evaluated, its characteristic connectivities $\mathcal{I}^n_{qj}=\left\lbrace i \in \left\lbrace 1,...,I\right\rbrace: \tilde{N}_i(\tilde{\xi}^n_{qj})\neq 0\right\rbrace$ are derived accordingly. Construction of this basis occurs such that $\mathcal{C}^{r-1}$ continuity is preserved, see \autoref{fig:fligaconceptPBC}.

Assume we have such a periodic parent basis $\left\lbrace\tilde{N}_{i}\left(\tilde{\xi}\right)\right\rbrace^I_{i=1}$ and ascending positions of the floating regulation points  $h^n_{ij} \in \left[ a^n, b^n \right]$ for given $j$. (Due to the float regulation point updates, $h^n_{ij} \notin \left[ a^n, b^n \right]$ is possible and then we shift $h^n_{ij}$ by a multiple of $b^n-a^n$ so as to enter this interval.) It is important to note that if we evaluate the floating regulation points $h^{n,q}_{ij}$ as connected to a parent quadrature point $\tilde{\xi}^n_{qj}$, we cannot set $h^{n,q}_{ij}=h^n_{ij}$ as this would cause severe spatial jumps of these points at the periodic boundary. Instead we choose $h^{n,q}_{ij}\in\left\lbrace h^n_{ij}+\lambda^{n,q}_{ij}\cdot(b^n-a^n)\right\rbrace$ with $\lambda^{n,q}_{ij} \in \left\lbrace -1,0,1 \right\rbrace$ such that
\begin{equation}
0<h^{n,q}_{(i+1)j} - h^{n,q}_{ij}<b^n-a^n
\end{equation}
while
\begin{equation}
a^n\leq\mathcal{G}^n_j\left(\tilde{\xi}^n_{qj}\right)<b^n
\end{equation}
We then follow the previously introduced B-Splines evaluation concepts for FLIGA.

%\nocite{*}
\bibliographystyle{elsarticle-harv}
\bibliography{Bib}
%%%%%%%%%%%%%%%%%%%%%%%%%%%%%%%%%%%%%%%%%%%%%%%%%%%%%%%%%%%%%%%%%%%%%%%%%%%%%%%%%%%%%%%%%%%%%%%%%%%%%%%%%%%%%%%%%%%%%%%

\end{document}